# Magnetoresistance in Hybrid Pt/CoFe$_2$O$_4$ Bilayers Controlled by Competing Spin Accumulation and Interfacial Chemical Reconstruction

Hari Babu Vasili*[†], Matheus Gamino[‡], Jaume Gàzquez*[‡], Florencio Sánchez[‡], Manuel Valvidares[†], Pierluigi Gargiani[†], Eric Pellegrin[†], and Josep Fontcuberta*[‡]

[†]ALBA Synchrotron Light Source, E-08290 Cerdanyola del Vallès, Barcelona, Catalonia, Spain

[‡]Institut de Ciència de Materials de Barcelona (ICMAB-CSIC), Campus UAB, 08193, Bellaterra, Catalonia, Spain

**ABSTRACT:**

Pure spin currents hold promises for an energy-friendlier spintronics. They can be generated by a flow of charge along a non-magnetic metal having a large spin-orbit coupling. It produces a spin accumulation at its surfaces, controllable by the magnetization of an adjacent ferromagnetic layer. Paramagnetic metals typically used are close to a ferromagnetic instability and thus magnetic proximity effects can contribute to the observed angular-dependent magnetoresistance (ADMR). As interface phenomena govern the spin conductance across the metal/ferromagnetic-insulator heterostructures, unraveling these distinct contributions is pivotal to full understanding of spin current conductance. We report here x-ray absorption and magnetic circular dichroism (XMCD) at Pt-M and (Co,Fe)-L absorption edges and atomically-resolved energy loss electron spectroscopy (EELS) data of $Pt/CoFe_2O_4$ bilayers where $CoFe_2O_4$ layers have been capped by Pt grown at different temperatures. It turns out that the ADMR differs dramatically, being either dominated by spin Hall magnetoresistance (SMR) associated to spin Hall effect or anisotropic magnetoresistance (AMR). The XMCD and EELS data indicate that the Pt layer grown at room temperature does not display any magnetic moment, whereas when grown at higher temperature it is magnetic due to interfacial Pt-(Co,Fe) alloying. These results allow disentangling spin accumulation from interfacial chemical reconstructions and for tailoring the angular dependent magnetoresistance.

## 1. Introduction

Proximity effects broadly refer to phenomena occurring when dissimilar systems, materials in the present case, are brought into close vicinity and results in a modification of their properties. Proximity effects may arise from a number or reasons. For instance, when perovskite oxides with different rotation or distortion patterns of the constitutive metal-oxygen octahedra are grown epitaxially on top of each other, the corresponding order parameters may not decay abruptly at the interface, but propagate to some extent into the neighboring layer inducing emerging structural, magnetic and electric properties.[1-2] Proximity effects may have quantum origin and reflect the fact that the wavefunctions cannot terminate abruptly at surfaces and/or interfaces and could thus leak across. A particularly well known example of major scientific and technological importance is the case of the superconducting wavefunctions that may extend into non-superconducting regions and allows supercurrents to flow across substantial lengths.[3-4] In the case of magnetism, magnetic proximity effects (MPE) can manifest themselves in a variety of forms. Of interest here, is the induction of magnetization in otherwise paramagnetic metals when brought in contact with ferromagnetic materials. In contrast to superconductors, this proximity-induced magnetic region is typically short range, of the order of only a few atomic distances, but in materials which are close to satisfying the Stoner criterion for ferromagnetism, such as Pd and Pt, it may to extend up to a few nanometers. In recent years MPEs are receiving a renewed attention due novel potential avenues for exploitation. For instance, MPE has been shown to be efficient in successfully bringing the Curie temperature of some topological insulators up to room temperature[5] or to induce spin polarization in graphene.[6-7] MPEs are also of relevance in the emerging field of dissipation-less spintronics due to the discovery that pure spin currents can be easily generated in metallic non-magnetic (NM) layers.[7-9] Spin currents can be obtained by spin pumping using the ferromagnetic resonance,[10] or thermal gradients,[11] or simply by driving a current along the NM layer.[9, 12] A typical device is formed by the NM metal having a large spin-orbit coupling, such as Pt, W, Ta, etc. deposited onto a ferromagnetic insulator (FMI). A charge flow within the NM layer produces a spin accumulation at the sample edges that may diffuse to the neighboring FMI layer (Figure 1b). The Pt/YIG bilayers, where YIG ($Y_3Fe_5O_{12}$) is a ferromagnetic insulator and Pt is a thin capping layer, is among the spin Hall systems that have attracted most attention and has thus been extensively investigated.[13-18] It was discovered that an angular-dependent magnetoresistance (ADMR) develops in the Pt layer which largely depends upon the direction of the magnetization of the underlying YIG layer. Nakayama[17] proposed that this ADMR originates from spin Hall effect and inverse Spin Hall effect in the non-magnetic Pt layer, arising from the reflected spin currents at the Pt/YIG interface, and henceforth named "spin Hall magnetoresistance (SMR).[17, 19] However, it has also been claimed that MPE may be at play, as magnetic moments can be induced in the Pt at the interface with the FMI (Figure 1a).[20-21] Their presence would contaminate not only the ADMR measurements but the spin conductance efficiency across NM/FMI interfaces in spin caloritronics or microwave-assisted spin pumping experiments, thus challenging a detailed understanding.

X-ray magnetic circular dichroism (XMCD) - being an atomically selective magnetization probe - has been used to assess the presence or absence of MPE at the Pt/YIG interfaces. Lu *et al.*[21] reported a clear XMCD at Pt-$L_{2,3}$ absorption edges in Pt/YIG thin films, thus concluding the prevalence of MPE. However, this observation didn't settle the question as Geprägs *et al.*[14] performed a similar experiment and found that the magnetic moment in Pt, if any, was within the experimental sensitivity (~ 0.003 $\mu_B$/Pt), thus concluding that MPE cannot be relevant; these authors further argued that the results reported in Lu *et al.* could be affected by a partial oxidation in the Pt film.[22] In any event, the nature of the interface is crucial for spin diffusion across it and, not surprisingly, thermal, mechanical or chemical surface treatments have been shown to modify the spin conductance across the Pt/YIG interfaces.[15-16, 23] SMR and longitudinal spin Seebeck effect measurements have also been used to evidence that in some Pt/YIG(111) bilayers, the magnetic field response appears to signal the presence of different magnetic textures at the bulk and surface of the YIG slab, probably related to the existence of some surface perpendicular magnetic anisotropy of extrinsic[23] or intrinsic origin.[24-25] Overall, these results indicate that the interface phenomena govern the spin conductance across the films.

Spinel oxides - such as $CoFe_2O_4$ (CFO),[26] $NiFe_2O_4$(NFO),[13] and $Fe_3O_4$ [13] -have also been used to perform similar ADMR experiments. In the case of CFO/Pt bilayers, the ADMR has been interpreted in terms of SMR[13] and it has been shown that the spin conductance across the Pt/CFO interface is sensitive to the magnetic texture of the interface, which can either be modulated by changing the film orientation[27] or its microstructure.[26] XMCD experiments at the Pt-$L_{2,3}$ and Pt-$M_3$ edges conducted on Pt/CFO bilayers grown on $SrTiO_3$(111) have determined that the Pt magnetic moment, if any, is below the detection limit (< 0.002 $\mu_B$/Pt).[28] X-ray resonant magnetic reflectivity (XRMR) measurements on the Pt/NFO bilayers have also confirmed that the magnetic moment of Pt is below the sensitivity limit of the technique (~0.02 $\mu_B$/Pt).[29] In any case, the XMCD or XRMR experiments referred above were used to assess the magnetic moment formation at Pt atoms, but these experiments did not give insight into the origin of any magnetic moment in Pt and dismissed any collateral effects that the growth of the metallic layer might have on the neighboring FMI layer (CFO, YIG or any other FMI). For instance, magnetic moment formation on Pt atoms could arise from genuine spin-dependent interfacial hybridization or, as emphasized recently[30] may result from an interfacial intermixing and non-well defined chemical Pt and CFO interfaces (Figure 1c).

Here, we explore in detail these issues with the purpose to shed light and obtain understanding the mechanisms that switches on/off magnetism at Pt at the Pt/CFO interface. We use CFO thin films because they are ferrimagnetic insulating oxides with large magnetization and, in contrast to the magnetically softer YIG, they exhibit a substantial magnetic anisotropy due to the presence of $Co^{2+}$ ions.[31] Therefore, the magnetic properties of CFO films are expected to be particularly sensitive to any electronic/atomic reconstruction that might occur at Pt/CFO interface. In order to obtain the clearest insight into interface-related phenomena, we have simultaneously grown two epitaxial CFO films on $MgAl_2O_4$(001) substrates and subsequently covered them with sputtered Pt(4nm) layers using different conditions. Note that, the growth conditions for the CFO films were identical while the Pt layers were grown at different temperatures. Although Pt is commonly grown by sputtering at nominally room temperature, it has been recently reported that even in that case, an unknown layer may exist at the interface between the insulator and a capping Pt layer.[32] Therefore, to exacerbate potential differences and to get a clearer insight by varying a single experimental growth condition, we chose to grow one Pt layer either at room temperature (sample RT) and the other at a high temperature of 400 ºC (sample HT). Magnetic measurements indicate that distinct thermal treatments produce some modification of the macroscopic magnetization of CFO films; however, dramatic changes are observed in their ADMR response. X-ray absorption spectroscopy and XMCD depth-sensitive measurements demonstrate that the growth conditions of Pt determine not simply the appearance of a magnetic moment in the Pt layer, but also a rich electronic and magnetic reconstruction at the CFO interface that governs the magnetic and magneto-transport properties of the bilayer.

High-resolution scanning transmission electron microscopy (STEM) data provide conclusive evidence of remarkable thermally-induced atomic reconstructions and Pt(Co,Fe) alloying at Pt/CFO interface (Figure 1c). Finally, by exploiting the ADMR, found to be larger in the HT than in the RT sample, and the substantial magnetic remanence of CFO films, we demonstrate that the magnetic state of the CFO films, written using suitable magnetic fields, can be inferred (read) by using a simple resistivity measurement of the neighboring Pt layer.

## 2. Experimental Method

CFO thin films of about 24 nm thickness were grown on (001)-oriented $MgAl_2O_4$ (MAO) substrates (5 x 5 mm$^2$). The films were grown from a CFO stoichiometric target by pulsed laser deposition (PLD) using a KrF laser ($\lambda$ = 248 nm) with fluence of 1.5 J/cm$^2$ and a repetition rate of 5 Hz at 450$^o$C in an oxygen pressure of 0.1 mbar. The typical surface roughness of the films grown under these conditions is of about ~0.13 nm (1 μm × 1 μm area scans) (Supplementary Information SI-1). The CFO films, grown in the same run, were further capped with a dc-sputtered Pt layer either grown *in-situ* at room temperature (sample RT) or at a high temperature of 400 ºC (sample HT). The thicknesses of the CFO films ($t_{CFO}$ = 24 nm, 40 nm) and Pt layers ($t_{Pt}$ = 4 nm) were deduced from the growth rate calibration obtained by x-ray reflectometry (Supplementary Information SI-1). The θ-2θ x-ray diffraction (XRD) scans confirmed that the CFO films are fully (00*l*)-textured and epitaxial (cube-on-cube) without any trace of spurious phases (Supplementary Information SI-1). The Pt layer grown at HT is also epitaxial, whereas when grown at RT it appears granular although some grains are epitaxial (Supplementary Information SI-1).

The microstructure of the two bilayers was examined by a JEOL ARM 200CF scanning transmission electron microscope with a cold field emission source operated at 200 kV and equipped with a CEOS aberration corrected and a Quantum electron energy loss spectrometer (EELS). Specimens for the STEM experiments were prepared by conventional thinning, grinding, dimpling, and Ar ion milling.

Longitudinal resistance and magnetoresistance on the Pt/CFO//MAO samples were measured by using four in-line silver paint strips as shown in Figure 2a. Spin Hall measurements were performed at room temperature (300 K) under an applied magnetic field of up to 90 kOe by using a Physical Property Measurement System (PPMS) equipped with a rotating sample holder. The room-temperature longitudinal base resistance ($R_L$) and resistivity ($\rho$) of the HT and RT Pt films ($t_{Pt}$=4 nm) are ~202 $\Omega$ ($\rho$ = 40.4 $\mu\Omega$-cm) and ~264 $\Omega$ ($\rho$ = 52.8 $\mu\Omega$-cm), respectively. These resistivity values (of the 4 nm Pt layer) are similar to those reported in the previous works.[13, 16, 20, 26-27, 33-34] In our experiments, the sample plane was defined by the current ($j_c$) flowing along the ***j***-direction, the ***t***-axis as the in-plane transverse direction, and the out-of-plane direction (***n***) (perpendicular to both ***j***- and ***t***-axes). Angular-dependent longitudinal magnetoresistance (ADMR) measurements were obtained by performing $\beta$- and $\gamma$- angular scans with the magnetic field (***H*** = 90 kOe) rotating either in (***n***, ***t***)- or (***n***, ***j***)-planes, respectively. The origin of the angular scans is defined by the ***n***-axis, as illustrated in Figure 2a. Remanent magnetoresistance ($R_{rem}(\beta)$) was obtained by applying the maximum field (90 kOe) at a given $\beta$ angle and zeroing $H$ before measuring the film resistance. The resistivity of the bare CFO film (24 nm) is of about $8.5 \times 10^5$ $\Omega$cm.

Magnetization measurements were carried out using a superconducting quantum interference device (SQUID) magnetometer (MPMS system of Quantum Design) with the field applied for either in-plane (***H//t***) or out-of-plane (***H//n***) directions at 300 K. Background signal subtraction procedure is described in Supplementary Information SI-1.

X-ray absorption spectroscopy (XAS) and XMCD measurements were carried out at 300 K at the BOREAS beamline of the ALBA synchrotron light source.[35] The XMCD spectra were obtained by taking the difference between the XAS of right ($\sigma^+$) and left ($\sigma^-$) circularly polarized light. We have used 70 % and 100 % degree of the circularly polarized light for the Pt-M and (Fe,Co)-L edges, respectively. In the XMCD measurements, the samples were magnetized by applying a magnetic field of 60 kOe in the beam direction for either in-plane (grazing incidence (20º from the film plane)) or out-of-plane (normal incidence) directions. The spectra were measured in the total electron yield (TEY) mode. To discriminate between the surface and bulk XMCD response, some spectra were also taken at normal incidence in both TEY and Fluorescence yield (FY) modes.

A complete set of measurements have been performed on a similarly prepared Pt/CFO(40 nm) samples, obtaining fully consistent results (Supplementary Information SI-2).

## 3. Results and Discussion

### 3.1. Magnetization and angular dependent magnetoresistance

Figures 2b,c show the room temperature M *vs* H hysteresis loops of the RT and HT Pt(4 nm)/CFO(24 nm)//MAO(001) samples (see Supplementary Information SI-1 for morphological, structural and magnetic characterization), respectively, obtained applying the magnetic field along in-plane (***t***-axis) and out-of-plane (***n***-axis) directions (geometry is sketched in Figure 2a). These magnetization loops indicate that: *i*) both samples have an in-plane easy magnetization direction (EMD) which results in a larger remanent magnetization ($M_{rem}$) for ***H//t*** than the ***H//n***;

*ii*) the high-field in-plane magnetization (***H***//***t***) of the RT sample is ~ 340 *emu/cm³* (2.69 $\mu_B$/f.u.) which is somewhat smaller than spin-only optimal value (~ 376 *emu/cm³*) but similar to that reported for CFO films of similar thickness,[36-38] in contrast, the in-plane magnetization (***H***//***t***) of the HT sample exhibits a significantly larger value (about 470 *emu/cm³* or 3.73 $\mu_B$/f.u.); *iii*) the ratio between the remanent magnetization [$M_{rem}$(***H***//***t***)/$M_{rem}$(***H***//***n***)], i.e the ratio between remanence for (***H***//***t***) and (***H***//***n***) at after magnetization the samples at 70 kOe is of about 12.4 and 6.2 for HT and RT samples, respectively; and finally *iv*) the coercive field (***H***//***t***) of the RT sample (~ 5 kOe) is slightly larger than that of the HT counterpart. We emphasize that these differences cannot be related to differences in the bulk properties of the CFO films (such as cationic inversion,[39] oxygen vacancies, *etc.*) because both films were grown simultaneously at 450 °C (before Pt deposition). Therefore, the difference magnetization behavior of HT and RT Pt/CFO samples must be related to the distinct conditions used for the growth of Pt on top of the CFO films.

More dramatic differences are observed in the longitudinal magnetoresistance (MR) data shown in Figure 3, where we plot the room-temperature MR $\left[ MR = \dfrac{R_L(H) - R_L(H=0)}{R_L(H=0)} \right]$ of the RT and HT samples recorded with ***H***//***j***. RT and HT samples display a radically different MR differing in both sign and magnitude. Indeed, the RT sample exhibits a small positive monotonic MR (~$2 \times 10^{-4}$ at 90 kOe) which is in good agreement with the reported Lorentz magnetoresistance values of the Pt films.[13, 17, 26-27, 40] In contrast, the HT sample has a negative non-monotonic MR which is larger in magnitude (by a factor of 37) than its RT counterpart. In addition, the MR shows a coercive field of ~3.9 kOe [Figure 3; see also Supplementary Information SI-1], which is close to the coercivity of the magnetization loops (***H***//***j***) of the HT sample (see Figure 2c). These strikingly different MR values reflect that the growth at 400 ºC of the Pt layer has induced critical changes in its electrical properties.

The ADMR for either β- or γ-rotation where the magnetization rotates in the (***n***, ***t***)- or (***n***, ***j***)- planes, respectively, is given by

$$\frac{\Delta R_L(\beta,\gamma)}{R_L} = \frac{\left\{ R_{\beta,\gamma}^{H(9T)} - R_{\beta,\gamma=90°}^{H(9T)} \right\}}{R_L} \qquad (1)$$

Figure 4a,b show the ADMR curves obtained by β- and γ-rotations, respectively, of the HT sample while Figure 4c,d contain the corresponding data for the RT sample. All data were recorded at 300 K using an applied field of 90 kOe. Both films display similar $\cos^2(\gamma,\beta)$ dependences in their ADMR response although its amplitude is much larger in HT than in RT.

According to the theoretical predictions of SMR, the longitudinal resistivity ($\rho_L$) is given by $\rho_L = \rho_0 + \rho_1(1-m_t^2)$ where $\rho_0$ is the baseline resistivity and $\rho_1$ is proportional to the SMR amplitude; $m$ ($m_j$, $m_t$, $m_n$) = $M/M_s$ are the direction cosines of the magnetization $M$ along the corresponding orthogonal directions, and $M_s$ is the saturation magnetization. Therefore, one would expect a $\cos^2\beta$ behavior in the ADMR curves (for β-rotation) as it is indeed observed in Figure 4a,c. For the RT grown sample (see Figure 4c) the amplitude of $\cos^2\beta$ oscillation is ~ $4.5 \times 10^{-4}$ which is in good agreement with the reported SMR values in similar Pt/FMI bilayers[13] suggesting its prevalence in RT Pt/CFO samples. On the other hand, the amplitude of the $\cos^2\beta$ oscillation of the HT sample is ~ $26.7 \times 10^{-4}$ which is significantly larger (about a factor 6) than its RT counterpart. This enhancement evidences that in the high-temperature processed sample other possible physical mechanisms may contribute to the observed ADMR.

We now turn to discuss the γ-rotation ADMR measurements where the response is more likely to arise, although not exclusively, from the combined effects of anisotropic magnetoresistance, henceforth AMR, ($\sim \sin^2\gamma$) and the Lorentz magnetoresistance (LMR) ($\sim \cos^2\gamma$), *i.e.* $R_L(\gamma) \approx A \sin^2\gamma + B \cos^2\gamma$, where A and B are constants. For a nearly saturated sample, during rotation along γ, $m_t \sim 0$ and therefore, SMR does not contribute to the ADMR. Data in Figure 4d show that the angular dependence of the $\frac{\Delta R_L}{R_L}$ of the RT grown Pt/CFO sample oscillates with a modest amplitude of $\sim 9 \times 10^{-5}$, that may result from the presence of a minor AMR+LMR contribution or from the fact that the magnetization of CFO films cannot be completely saturated even at 90 kOe. Indeed, it is known that in CFO and other spinels thin films, antiphase boundaries caused by the stacking faults during the growth of the films, prevent saturation.[41] In contrast, in the HT grown Pt/CFO sample, the amplitude of the angular dependence is much larger ($\sim 4.\ 5 \times 10^{-3}$) (see Figure 4b). The magnetization data discussed above, showed that the magnetization ratio $\frac{M_{70\text{kOe}}(H//t)}{M_{70\text{kOe}}(H//n)}$ of HT and RT samples, were of slightly different (1.7 and 1.4, respectively). Therefore, the astonishing increment in the ADMR in HT cannot be attributed to the differences on CFO magnetization but it should have a different origin. As this corollary is fully consistent with the conclusions derived from the β-rotation experiments described above, we conclude that an important interface-related magnetic modification has occurred in the HT sample. The simplest ones could be that in the HT sample the Pt layer has some magnetic character, either induced by MPE[20, 42] or due to some Pt/CFO chemical interdiffusion. This conclusion is confirmed by transverse magnetoresistance measurements. Results show a pronounced anomalous Hall effect in the HT sample which is absent in the RT sample (see Figure S8, Supplementary Information SI-1). In the following, we disentangle the contributions of either proximity effects or intermixing effects.

## 3.2. X-ray absorption and magnetic circular dichroism

Element-specific XMCD has been widely used to explore the presence of magnetic moments in non-magnetic metals, such as Pt, grown on ferromagnetic insulating oxides (like YIG or CFO).[14, 21, 28] Consequently, we have performed the XMCD experiments at Pt-$M_3$ and (Fe,Co)-$L_{2,3}$ x-ray absorption edges on RT and HT samples to elucidate the presence of magnetic moments on the Pt as well as any electronic and magnetic reconstruction occurring at the Fe/Co ion sites of the underlying CFO thin films. In the following, we discuss first the results of Pt-$M_3$ absorption edges and next we will describe the (Fe,Co)-$L_{2,3}$ edges.

The Pt x-ray absorption near-edge structure (XANES) spectra of $M_{2,3}$ edges (3p→5d) provide the same information as the commonly used $L_{2,3}$ edges (2p→5d). From the reported Pt $M_{2,3}$ XANES spectra, it is known that the $M_3$ edge exhibits sharp spike at the absorption threshold (so called "white line (WL)") which is less pronounced or even absent at the $M_2$ edge.[43] Therefore, we restrict our discussion only to the $M_3$ edge ($3p_{3/2} \rightarrow 5d_{5/2}$). Figure 5a,b show the room-temperature XAS and XMCD of Pt-$M_3$ absorption edges for the RT and HT samples, respectively, at normal incidence. We first note that the Pt XANES spectra of HT and RT samples look similar to each other, with a minor increase in the ratio of WL intensities (1.12: 1.20) from HT to RT samples (see the Supplementary Information SI-1). The above Pt-$M_3$ WL intensities (and further scaling up the $M_3$ intensities to $L_3$ edges) are very much similar to the reported Pt/CFO of Pt grown at room temperature [Ref. 44] and 400°C [Ref. 28]. Overall, our data unambiguously revealing that the Pt of HT sample is metallic while the RT sample probably has some partial oxygen adsorption. This Pt oxide phase reduces with the temperature, therefore, the HT grown Pt becomes metallic.

Although the RT sample is magnetic, but no appreciable XMCD was observed at the Pt-$M_3$ edge (see Figure 5a) which indicates the absence of MPE in the Pt layer grown at room-temperature. This is in contrast with the analogous measurement on the HT sample shown in Figure 5b that evidences an XMCD of about 4% in units of the white line intensity. To determine the corresponding magnetic moment we follow the calibration procedure reported in Ref. [28] where the spectra were normalized to the average white line peak intensity. With this, the $M_3$-XMCD yields an approximate calibration of 0.06 $\mu_B$/Pt per 1% signal,[28] which indicates that the XMCD of the HT sample (~ 4 %) correspond to ~ 0.24 $\mu_B$/Pt atom. Such value is similar to those obtained in for example Co/Pt interfaces.[28] Note that, the above HT sample shows a similar amount of dichroism both in-plane and out-of-plane magnetization.

However, the mere observation of an induced Pt magnetic moment in the HT grown Pt/CFO sample doesn't provide any hint on the underlying physical mechanism behind its formation. In other words, it cannot discriminate between being either the result from a proximity effect or any interdiffusion at the Pt/CFO interface. In order to get deeper understanding onto the underlying physical mechanism and aiming to discriminate between proximity effects or any interdiffusion at the Pt/CFO interface, we have performed XAS and XMCD experiments at (Fe,Co)-$L_{2,3}$ absorption edges of the CFO in both Pt/CFO bilayers and complementary STEM-EELS atomic mapping.

The TEY signals of (Fe,Co)-$L_{2,3}$ XAS and XMCD of the RT sample at grazing incidence [Figure 6a,b for Fe-$L_{2,3}$ and Co-$L_{2,3}$ edges, respectively] show the spectral features typically observed in the CFO films:[28, 45-48] the Fe-L$_3$ XMCD displays one positive (~709 eV) and two negative peaks (~708, ~710 eV) corresponding to their tetrahedral ($T_d$) and octahedral ($O_h$) site occupancies, respectively, whose spins are known to be antiparallel to each other [Figure 6a]. These features are similar to those observed in the isostructural γ-Fe$_2$O$_3$ oxide,[49-50] indicating that iron ions of the CFO film are in $3d^5$-Fe$^{3+}$ state. Our calculated Fe$^{3+}$ XMCD spectra using ligand field multiplet calculations[51] are in very good agreement with the above experimental spectra [see the inset of Figure 6a]. On the other hand, the shape of the Co-L$_3$ XMCD signal is that of Co$^{2+}$ ($3d^7$) in spinel oxides [Figure 6b]. The calculated Co-L$_3$ XMCD spectra - see inset Figure 6b- indicates a prevailing Co$^{2+}$($O_h$ and $T_d$ symmetry) valence state.

We summarize the above spectroscopic measurements as follows: *i*) (Fe,Co)-L XMCD signals are similar to those commonly reported for (uncapped) CFO films and *ii*) the absence of Pt magnetic moment. We conclude that when depositing Pt on CFO//MAO(001) at room-temperature, Pt neither displays any magnetic proximity effect nor diffuses into or reacts with the CFO film underneath.

In contrast, the (Fe,Co)-L$_3$ XMCD spectra of the HT Pt/CFO sample are remarkably different. First, the Fe-L$_3$ XMCD displays two negative peaks (large peak at ~708 eV and a small one at ~710 eV) and one very small positive peak at ~709 eV [see Figure 6c]. The peaks at ~709 eV and ~710 eV are analogous to the spectral features as observed in the RT sample (Figure 6a) which reflects a partial presence of the Fe$^{3+}$ on the $O_h$ and $T_d$ sites. However, the larger peak at ~708 eV is indeed broader than the Fe$^{3+}$ peaks and hints to a Fe metallic content (Fe$^0$) in the sample. To make more evident this metallic Fe$^0$ contribution, we have subtracted (properly weighted (~30 %)) the Fe$^{3+}$ XMCD of the RT sample from that of the HT sample as shown in Figure 6c (inset). Second, the shape of the Co-L$_3$ XMCD (see Figure 6d) closely resembles that of a reduced metallic-like Co signal ($Co^0$) with very small features reminiscent of the presence of $Co^{2+}$. After subtracting the ~5% of RT $Co^{2+}$ XMCD from the HT XMCD, we get the metal-only cobalt content (inset of Figure 6d). The presence of these metallic Fe/Co signals in the TEY spectra indicate the formation of an interfacial Pt-(Co,Fe) alloy. As a consequence, the Pt becomes magnetic through the $3d$-$5d$ exchange and as aforementioned it exhibits the magnetic moment $m$(Pt) comparable to the Co-Pt alloys or Co/Pt thin films.[28]

Electronic reconstructions at the CFO/Pt interface should imply a change of the magnetic moment of the Fe and Co ions. We have determined the spin ($m_s$) and orbital ($m_l$) magnetic moments of (Fe,Co) ions/metals and so the magnetization of the CFO film/Pt-(Co,Fe) alloy by applying the sum rules to the integrated intensities of total XAS and XMCD spectra.[52-54] It turns out (see Supplementary Information SI-3 for details) that Fe at HT sample display a bulk-like total magnetic moment of ~2 $\mu_B$, which is consistent with the above mentioned presence of $Fe^0$ metal. Overall, the total magnetic moment ($m_{tot}$) as determined by the XMCD analysis of Pt-(Co,Fe) alloy [3.60(5) $\mu_B$] is significantly larger than that of the RT sample [2.40(5) $\mu_B$]. The $m_{tot}$ of RT sample is in good agreement with our aforementioned SQUID magnetization data (2.69 $\mu_B$) and within the range of XMCD-extracted values reported for CFO thin films (2-3 $\mu_B$) at 300 K.[36-38]

The formation of a ferromagnetic Pt-(Co,Fe) alloy at the CFO/Pt interface explains the observed larger magnetization (Figure 2c), the larger magnetoresistance (Figure 3), and the enhanced ADMR (Figure 4a,b) in the HT Pt/CFO//MAO sample. From the spectral fingerprint in the XMCD spectra (see Figure 6c,d), it thus follows that Pt does not get magnetic by proximity effect but rather from an interdiffusion / interface-alloying process.

Having used the XMCD to demonstrate the appearance of a magnetic moment in Pt and a concomitant reduction of $Fe^{(3-\delta)+}$ and $Co^{(2-\delta^*)+}$ ions eventually forming a ferromagnetic metallic $(Co_xFe_y)Pt_z$ alloy at the Pt/CFO interface, it remains to be confirmed that the observed changes are limited only to the interface. To this purpose we have measured the XMCD at (Fe,Co)-$L_{2,3}$ edges in both the TEY and FY modes. TEY mode is sensitive to CFO film surface as the escaping depth ($\lambda_e$) of photoelectrons at the used energy is about 2-5 nm (for e.g. $\lambda_e$ is 5nm for $Fe_3O_4$, Ref. 55). FY basically probes possibly a full depth of the film due to the larger $\lambda_e$ of fluorescence photons (of the order of several tens of nanometers). Therefore, comparison of TEY and FY data could enable to obtain a depth-profile analysis of the element-specific magnetic response. We used normal incidence because the incoming photons can penetrate deeper -so larger $\lambda_e$ - along the surface normal direction. In Figure 7a-d, we show the XMCD of (Fe,Co)-$L_{2,3}$ edges of the HT sample recorded by TEY and FY modes, in the upper and lower panels, respectively. The sketches (left side) in the Figure 7 illustrate the x-rays penetration into the film and emphasize that in TEY experiments probe electrons escaping from about $\approx \lambda_e$ while FY probes the full sample thickness.

The TEY (Fe,Co)-XMCD data (Figure 7a,b) of the HT sample recorded at normal incidence are similar to those recorded at grazing incidence (see Figure 6c,d). However, the FY spectra mode are definitely different (Figure 7c,d). Most obvious difference is that the typical signatures of ($Fe^{3+}$,$Co^{2+}$) valence state in the CFO are well apparent in the FY signal (see Figure 7c,d) but metallic-like (Fe,Co) is observed in the TEY signal (see Figure 7a,b). It appears that the reduced $Fe^{3+}/Co^{2+}$ into metallic $Fe^0/Co^0$ are concentrated at the Pt/CFO interface whereas deep into the CFO film, $Co^{2+}$ and $Fe^{3+}$ species, as expected in the CFO structure, remain prevalent. On the other hand, the TEY and FY data of the RT sample has no difference in their spectral features, which are characteristic of $Co^{2+}$ and $Fe^{3+}$ species, indicating the absence of any interdiffusion at the interface.

A complete set of similar SQUID and XMCD experiments have been performed on a Pt(4 nm)/CFO(40 nm)//MAO(001) where the thickness of CFO has been roughly doubled. Results are fully consistent (Supplementary Information SI-2).

### 3.3. High resolution electron microscopy and Electron energy loss spectroscopy atomic mapping

A detailed study of the Pt/CFO interface was done by means of the STEM experiments. When acquired with a high-angle annular detector, the STEM images present the so-called Z-contrast (being Z the atomic number).[56] The darkest contrast in Figure 8a,b corresponds to the MAO, followed by that of the CFO film. The brightest contrast is due to the heaviest Pt atoms, which allow for probing any interdiffusion of it within the CFO layer. The images show a sharper Pt/CFO interface when the Pt is deposited at room temperature (see Figure 8a,c) than when it is deposited at high temperature (see Figure 8b,d). Besides, two important structural changes in the HT are worth noticing. First, a mesa-like surface reconstruction with well-defined [101] CFO mesa-edges at 45°, reflecting a CFO surface reconstruction during HT growth of Pt and the subsequent formation of the lowest surface energy {111} pyramidal facets at the CFO surface.[57] Second, an evident Pt epitaxial texture is best observed in the HT sample (see Figure 8c,d).

Alongside, we used the combination of STEM and EELS to probe the possible chemical segregation of Fe and Co within the Pt layer. Figure 8e-g and h-j show the O, Fe and Co relative composition maps of the RT and HT samples, respectively. The averaged profiles of the relative composition maps (Figure 8e-g and h-j right panels) show neither depth-wise variations of the CFO film composition in the RT sample, nor traces of Fe or Co within the Pt layer.

On the other hand, as shown by the averaged relative composition profiles, the HT sample shows not only Fe enrichment and the Co impoverishment near the Pt/CFO interface, but also the presence of Fe and Co signals inside the Pt layer. A further evidence comes from the analysis of the EEL spectra collected at the Pt region, see Figure 9a, in which the spectrum from the HT sample (in black) shows both the Fe and the Co signals, whereas they are absent in the spectrum obtained from the RT sample (in red). It is also worth noting that none of the spectra show evidence of oxygen (the O-K edge appears around 530 eV) within the Pt layer.

The EEL spectra not only provide information about the chemistry of the material but also on its electronic properties. Two Fe and Co $L$-edges are compared in Figure 9b,c taken from the HT sample, one collected within the CFO layer (in black) and one from within the Pt layer (in red). The two set of signals are definitely different, being the most notorious the intensity ratio of Fe-$L_3$ and -$L_2$ white lines, known as $L_{23}$ ratio, which correlates with Fe and Co valences.[58-59] Indeed, the Fe and Co signals from within the Pt layer show distinctive metallic-like features, a lower $L_{23}$ ratio than their CFO counterparts and a shifting to lower energies of the Fe- and Co-$L_3$ and $L_2$ peaks.[60-61] These results confirm the XMCD anticipated conclusions, i.e. the formation of a Pt-(Fe,Co) alloy at the Pt/CFO interface during HT Pt deposition.

### 3.4. Electrical retrieving of the magnetic remanent state

The sensitivity of ADMR in the Pt layer with respect to the magnetization direction of the neighboring CFO film, combined with the substantial anisotropic magnetic remanence of CFO, offers the possibility to write a given magnetic state into the CFO film and a subsequent reading by using the magnetoresistance of the adjacent Pt layer. The simplest protocol is to magnetize the CFO film in two perpendicular directions $\boldsymbol{n}$ ($\beta$ = 0, 180º) and $\boldsymbol{t}$ ($\beta$ = 90º, 270º). After zeroeing the applied magnetic field remanent magnetizations $M_{rem}$ ($\beta$ = 0, 180º) and $M_{rem}$ ($\beta$ = 90º, 270º) are established and we measure and compare the corresponding longitudinal magnetoresistance (Figure 10 (sketches)).

Samples were magnetized, at room temperature, by applying the field of 90 kOe along $\boldsymbol{j}$ and we measured the resistance $R_{rem}(\beta) = R_L(H = 0, \beta(0, 90º))$. We define the relative change of remanent magnetoresistance as:

$$\frac{\Delta R_{rem}}{R_{rem}} = \frac{R_L\left[M_{rem}(\beta=0°)\right] - R_L\left[M_{rem}(\beta=90°)\right]}{R_L\left[M_{rem}(\beta=90°)\right]} \tag{2}$$

The $\Delta R_{rem}/R_{rem}(\beta=90º)$ values of RT and HT samples, calculated using Eq.(2) are plotted in Figure 10. We first note that the magnitude of the $\Delta R_{rem}/R_{rem}(\beta)$ oscillation is much larger in HT sample than for the RT counterpart. This is consistent with the ADMR differences observed in the $\beta$-scans for both the samples shown in Figure 4a,c. More important here is that for both samples $\Delta R_{rem}/R_{rem}(\beta=0, 180º)$ is larger than $\Delta R_{rem}/R_{rem}(\beta=90, 270º)$. This result is directly connected with the magnetization loops of the Pt/CoFe$_2$O$_4$ bilayer shown in Figure 2b,c, where the remanent magnetization ($M_{rem}$) was found to be larger for $\boldsymbol{H}//\boldsymbol{t}$ than for $\boldsymbol{H}//\boldsymbol{n}$, $i.e.$ $M_{rem}(\beta = 90º) > M_{rem}$ ($\beta$ = 0º). Therefore, for each film, the $M_{rem}$ dictates the amplitude of $\Delta R_{rem}/R_{rem}(\beta)$. Data in Figure 10 reveals that $R_L(M_{rem}(\beta = 0, 180º)) > R_L(M_{rem}(\beta = 90º, 270º))$ and thus a larger $M_{rem}$ is connected to the smaller $R_L$. Therefore, the ADMR of Pt allows inferring the remanent magnetization direction of the CFO film.

**4. Final remarks and conclusions**

In summary, the growth conditions of the Pt layer, namely the growth temperature, determine the properties of the Pt/CFO interface. In the present case, the Pt/CFO bilayers were grown on MAO(001) substrates, and as a consequence the CFO films are (001) textured. In general, this would correspond to CFO epitaxial films with exposed (001) surfaces. However, in spinel oxides, typically the {111} planes are the ones with smaller surface energy thus implying that the actual surface of the CFO film should display a competition of substrate-induced (001) planes and {111} nanofacets whose relative abundance is critically dependent on growth conditions, CFO thickness and subsequent thermal treatments.[57] The relative stability of Pt on (001)- and (111)-surfaces largely differs, as confirmed by recent experimental and first principle calculations of Pt/MgAl$_2$O$_4$.[62] It thus follows that CFO(111) interfaces should be less prone to chemical/electronic/magnetic instabilities. The absence of "proximity effects" in similarly grown Pt/CFO bilayers on (111)-oriented SrTiO$_3$ is consistent with this view.[28] Resulting from this surface instability, further stimulated by the high temperature deposition and low pressure used to grow the Pt capping layer, the CFO(001) layers close to the Pt/CFO interface are reduced and cation interdiffusion occurs which leads to the formation of a (Co,Fe)Pt alloy. Therefore, magnetism in Pt does not result from a genuine proximity effect but, at least partially, by the structural, electronic and magnetic interface reconstructions. The new magnetic phases formed during high temperature Pt capping, likely to be a metallic alloy, provide an additional contribution to the angular dependent magnetoresistance that overrules and masks the spin Hall magnetoresistance. The XMCD experiments at Fe, Co and Pt edges together with the EELS data reported here have been instrumental to unravel the unexpected magnetic reconstructions that may occur at interfaces between Pt and insulating ferromagnetic oxides. Whereas the interface reconstruction has been on-purpose exacerbated here by the use of a relatively high-temperature Pt deposition, interface alloying might be promoted by precise sputtering and oxide surface conditions even in nominally room-temperature growth. Hence, these results appear very relevant at the level of the selection of materials and their processing conditions for spintronic devices based on ferrimagnetic insulating oxides. Finally, the demonstration of the possibility of reading the magnetic state of an insulator via the magnetoresistance (Spin Hall magnetoresistance and/or anisotropic magnetoresistance) in a neighboring metallic layer, may open new perspectives challenged by the relatively small values of ADMR measured here.

**FIGURES:**

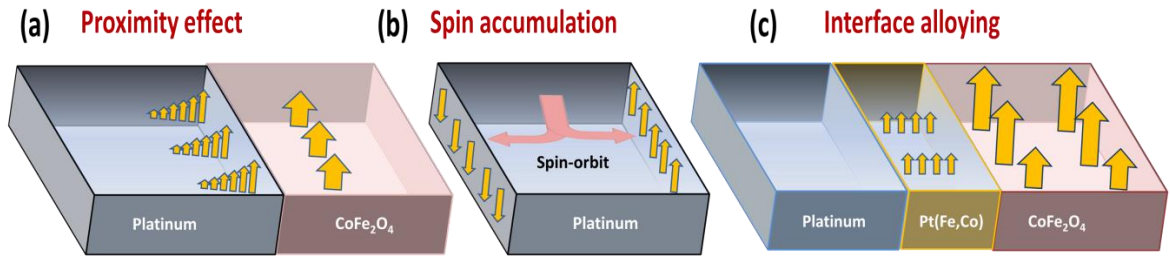

**Figure 1.** Sketch of magnetization distribution at the neighborhood of interfaces in paramagnetic-heavy metal/FMI heterostructures. (a) and (c) illustrate magnetic proximity effect and intermixing respectively. In presence of a charge flow, spin Hall effect promotes spin accumulation at interfaces, as illustrated in (b).

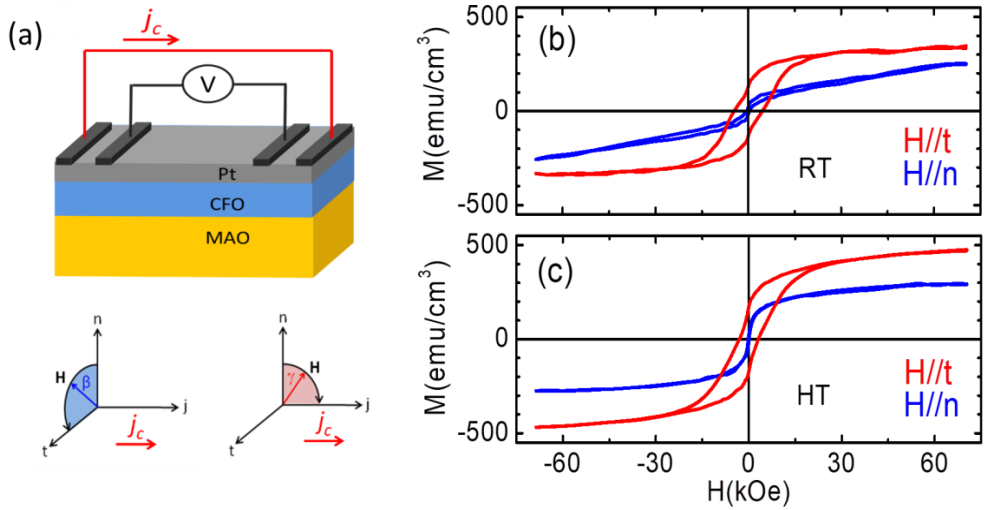

**Figure 2.** (a) Schematics of contact configuration for MR measurements and magnetic field rotation angles. The external magnetic field (*H*) is either applied in the (***n, t***) plane (β angle) or in the (***n, j***) plane (γ angle). In-plane and out-of-plane magnetization loops measured at 300K for Pt(4nm)/CFO(24nm) thin films deposited on MAO(001) substrate, with the Pt layer deposited at (b) room temperature (RT) and (c) at 400 °C (HT).

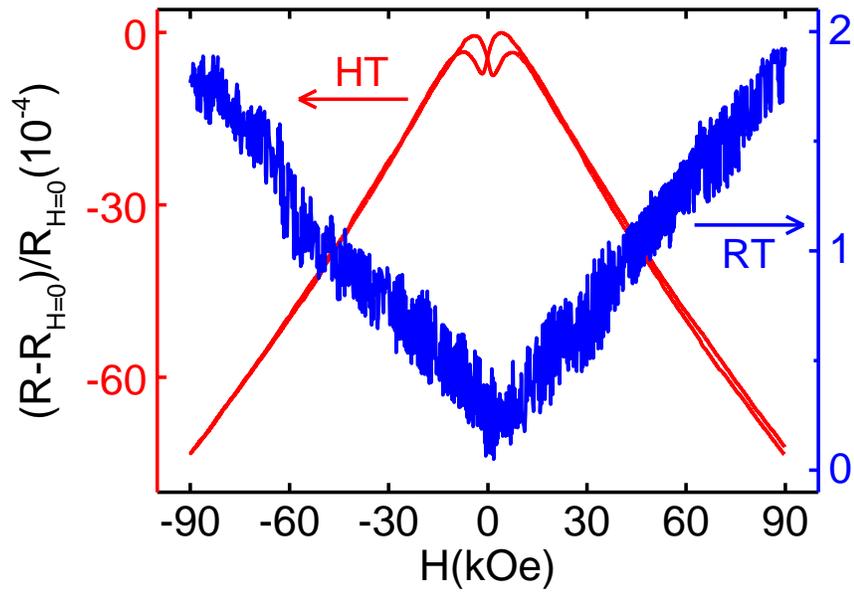

**Figure 3.** Room temperature MR of the RT (right axis) and HT (left axis) grown Pt/CFO//MAO(001) samples for ***H***//***j*** direction.

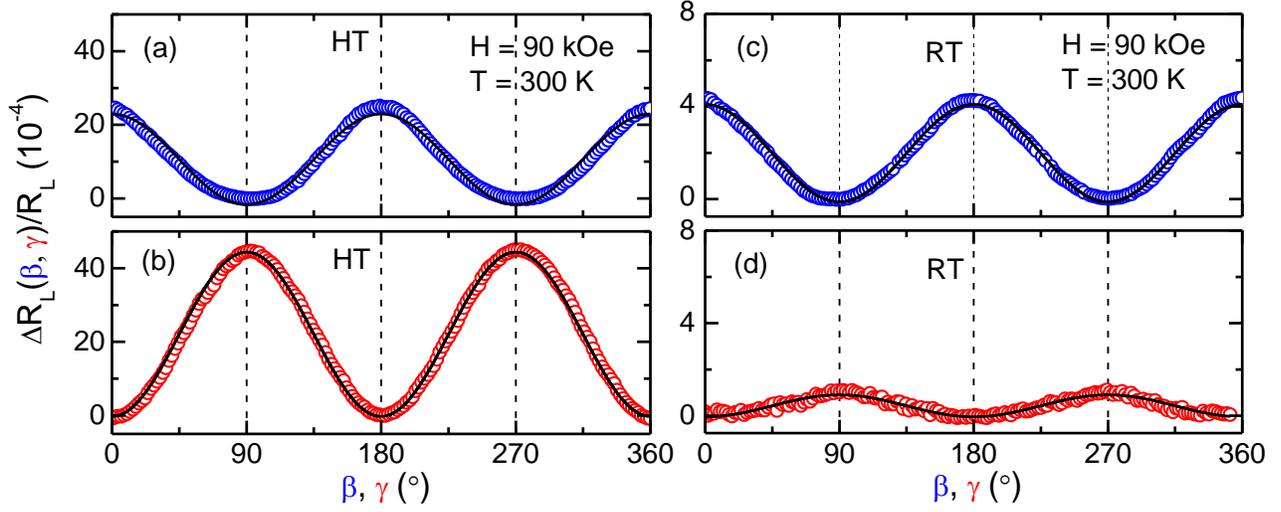

**Figure 4.** ADMR of (a,b) HT and (c,d) RT samples by applying the field of 90 kOe at β -(upper panels) and γ -(lower panels) angles, measured at 300 K. The solid lines are the best fits to the data using A sin²γ + B cos²γ (bottom panels) and A cos²β (top panels).

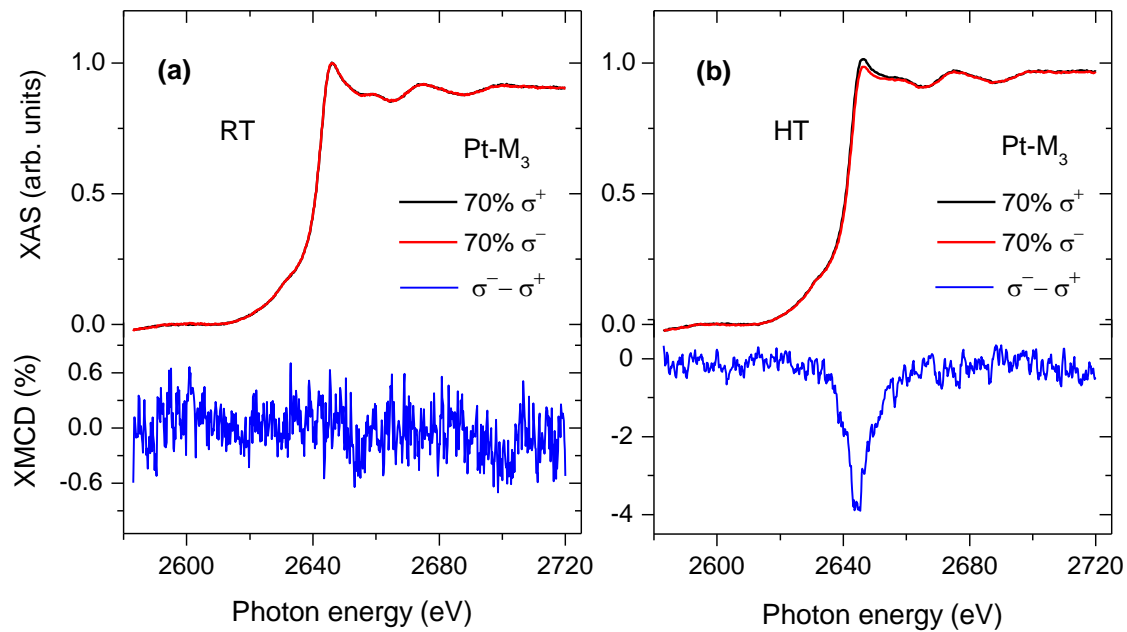

**Figure 5.** The Pt-M$_3$ XAS and XMCD of (a) RT and (b) HT grown Pt/CFO//MAO(001) samples, measured at 300 K. The spectra are normalized to unity at the white line peak intensity at 2646 eV.

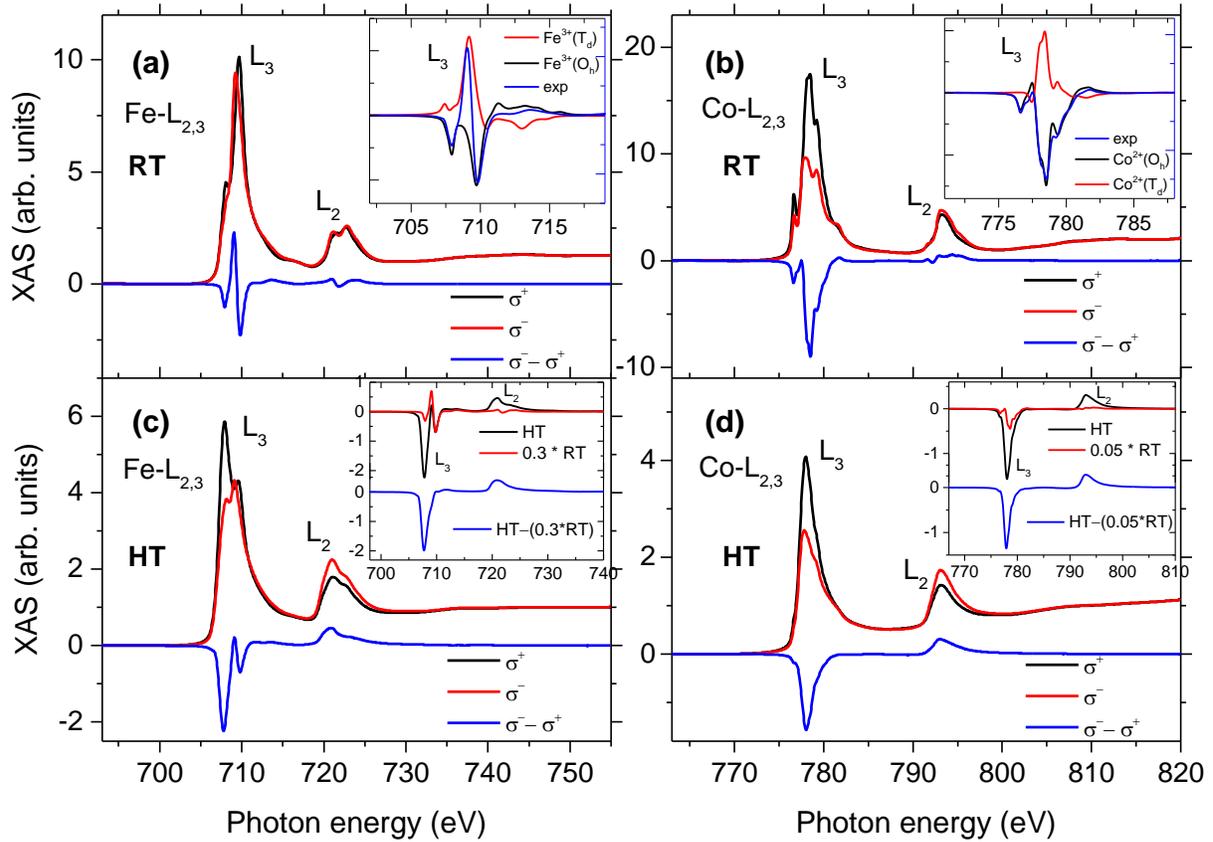

**Figure 6.** (a,c) Fe-$L_{2,3}$ and (b,d) Co-$L_{2,3}$ XAS and XMCD spectra of RT (upper panels) and HT (lower panels) Pt/CFO//MAO(001) samples, recorded in TEY mode at grazing incidence at 300 K. Insets (a,b) shows the calculated XMCD spectra for $Fe^{3+}(T_d,O_h)$ and $Co^{2+}(T_d,O_h)$ ions, respectively. Insets (top panels) in (c,d) show the Fe and Co edges respectively, of the HT samples (black lines) and the properly scaled (by a factor $\kappa = 0.3$ and $0.05$) corresponding spectra of RT samples (red lines). Insets (bottom panels) in (c,d) show the corresponding HT - $\kappa$RT difference spectra.

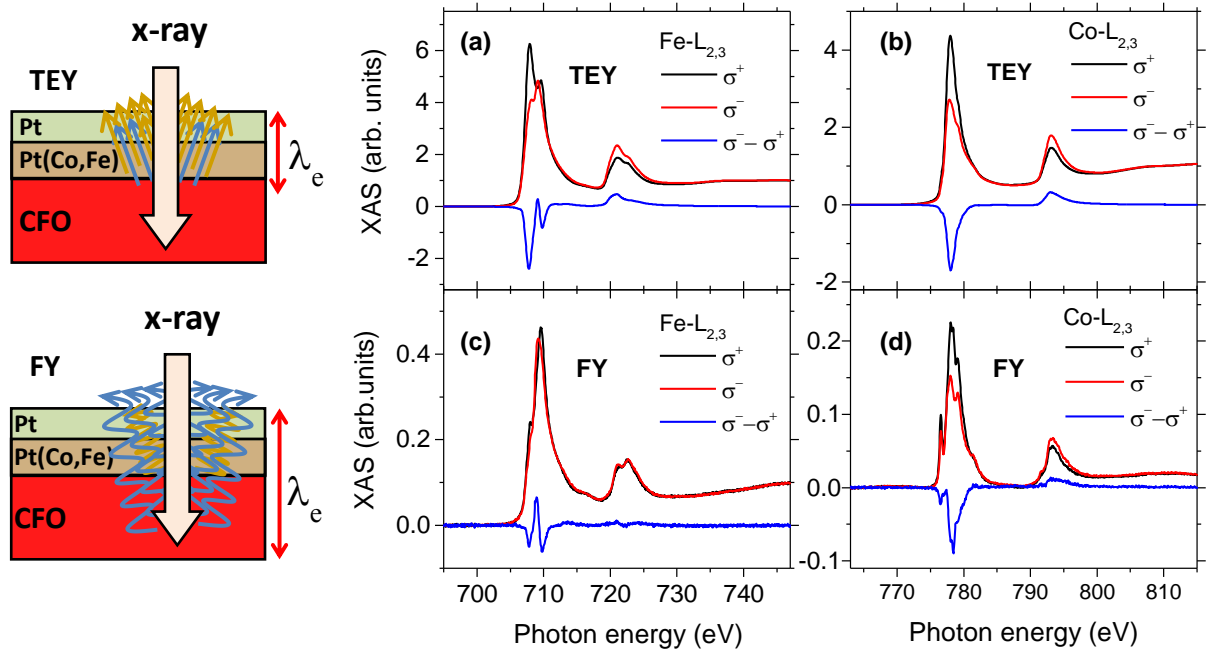

**Figure 7.** (a,c) Fe-$L_{2,3}$ and (b,d) Co-$L_{2,3}$ XAS and XMCD spectra of the HT Pt/CFO//MAO(001) sample, recorded using TEY (upper panels) and FY (lower panels) at normal incidence, at 300 K. The XMCD in the TEY signals is surface sensitive (see the left side upper sketch, where the $\lambda_e \sim$ 5 nm approx.) while the FY signals probe the full depth of the films (the left bottom sketch). Straight arrows and the curved arrows in the sketches are referred to the TEY and FY signals (electrons and photons, respectively), with different absorption cross section for different incoming photon helicities (blue and orange), respectively.

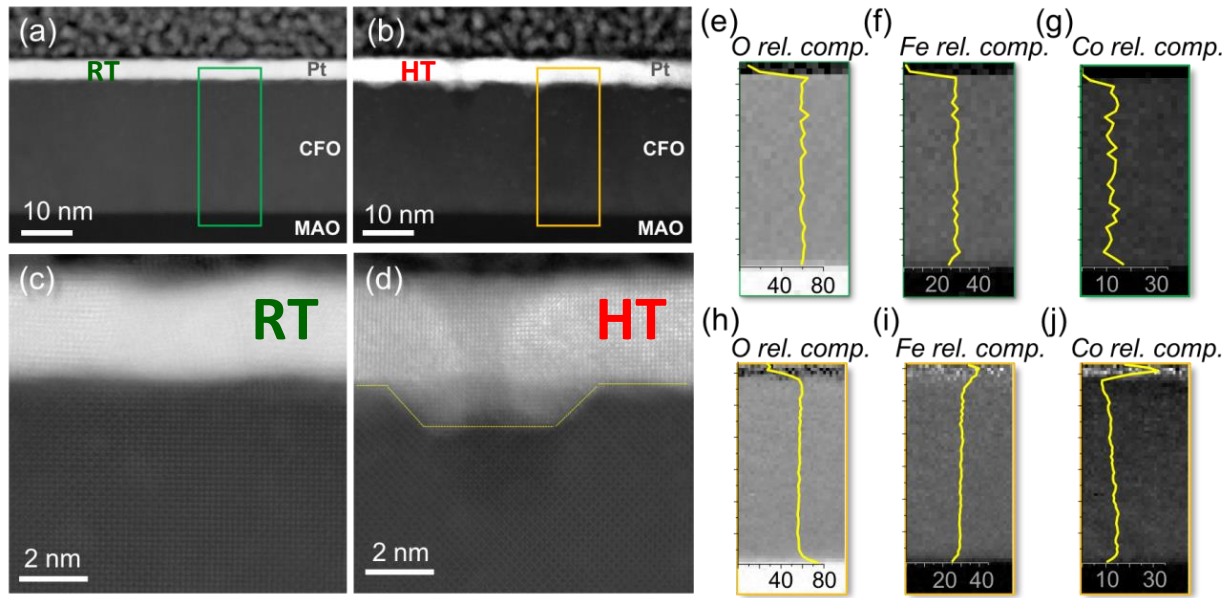

**Figure 8.** (a) and (b) are low magnification Z-contrast images of the Pt/CFO/MAO bilayers, viewed along the [100]MAO zone axis, of samples where the Pt was deposited at RT and HT, respectively. Pt and CFO layers are continuous over long lateral lengths. No secondary phases were observed. MAO and CFO epitaxial relationship is: MAO(100)[001] // CFO(100)[001]. (c) and (d) high-resolution Z-contrast images of the Pt/CFO interfaces with the Pt deposited at RT and at HT, respectively. The CFO film of the sample with Pt deposited at HT show a mesa-like surface with well-defined [101] CFO mesa-edges, which are marked with yellow dashed lines. The green and yellow rectangles mark the area where spectrum images were acquired, for which integration windows 30 eV wide were used after background subtraction using a power-law fit. (e-g) and (h-j) show the relative composition maps (in %) corresponding to the O K, Fe L and Co L edges from the samples with the Pt deposited at RT and with the Pt deposited at HT, respectively. Line traces show the corresponding depth-wise averaged profiles (in %).

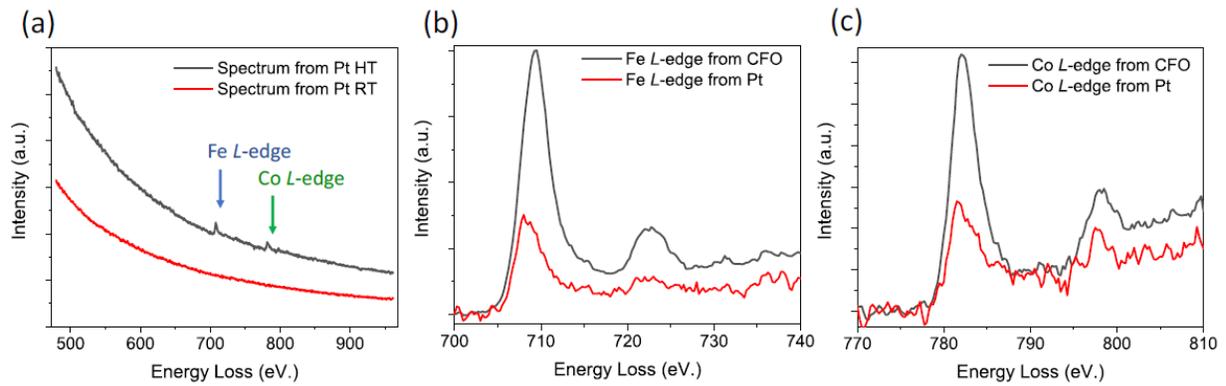

**Figure 9.** (a) EEL spectra collected at the Pt layer from the samples with Pt deposited at RT (in red) and at HT (in black). The blue and green arrows mark Fe and Co L-edges, respectively. (b) and (c) Fe- and Co-L edges collected at the CFO layer (in black) and at Pt layer (in red) from the HT sample, respectively.

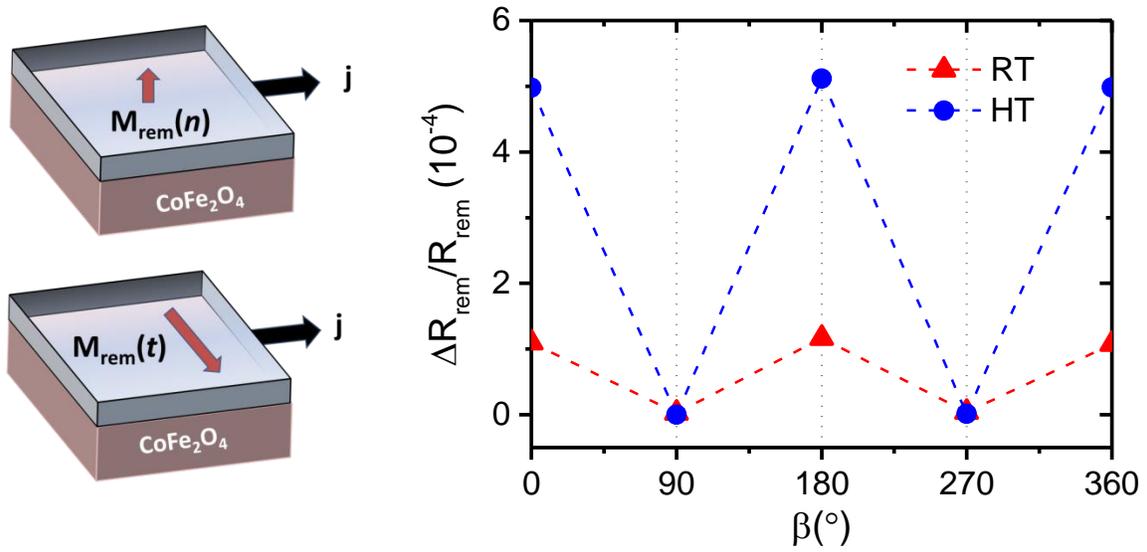

**Figure 10.** Left panel: sketches of measurements of remanent magnetoresistance. Right panel: Remanent magnetoresistance of Pt/CFO thin films at 300 K, measured at remanence after applying the magnetic field normal to the film (β = 0º, 180º) and parallel to the film (β = 90º, 270º). The dash lines are eye-guides. $\Delta R_{rem}$ is defined by difference between $R_L$(β = 0º) and $R_L$(β = 90º) measured at remanence normalizing by the resistance recorded at β = 90º. Solid circles (blue color) and triangles (red color) are referring to the HT and RT samples, respectively.

**ASSOCIATED CONTENT**

**Supporting Information**

Detailed supplementary information (SI) on (1) Structural, morphological, and magnetic properties of CFO thin films by XRR, XRD, STEM, AFM, and SQUID; and Pt-XAS analysis; (2) Magnetic properties of Pt/CFO//MAO before and after Pt growth at high temperature by SQUID and XMCD analysis; (3) Determination of atomic magnetic moments of HT and RT samples by Sum rule analysis.

**AUTHOR INFORMATION**

**Corresponding Author**

*E-mail: hvasili@cells.es (H.B.V.)

*E-mail: jgazqueza@gmail.com (J.G.)

*E-mail: fontcuberta@icmab.cat (J.F.)

**ORCID**

Hari Babu Vasili: 0000-0002-3356-6436

Jaume Gàzquez: 0000-0002-2561-328X

Josep Fontcuberta: 0000-0002-7955-2320

**Notes**

The authors declare no competing financial interest.

**ACKNOWLEDGMENT**

This work was supported by the Spanish Government by the MAT2014-56063-C2-1-R, MAT2017-85232-R and the Severo Ochoa SEV-2015-0496 Projects, the Generalitat de Catalunya (2014 SGR 734 Project). We acknowledge fruitful discussions with F. Casanova. M.G. thanks his fellowship from CNPq - Brazil. XAS and XMCD experiments were performed at the Boreas beam line of the Synchrotron Light Facility ALBA with the collaboration of ALBA staff. J.G acknowledges RyC contract (2012-11709). Electron microscopy observations carried out at the ICTS-CNME at UCM. Authors acknowledge the ICTS-CNME for offering access to their instruments and expertise.

**Table of Contents (TOC):**

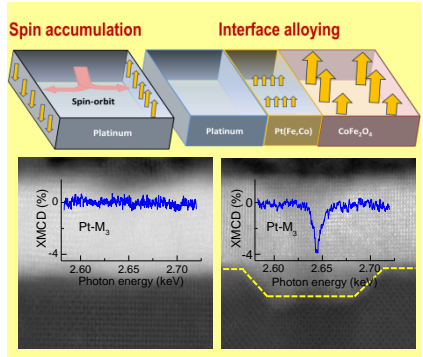

# Supporting Information

**Magnetoresistance in Hybrid Pt/CoFe$_2$O$_4$ Bilayers Controlled by Competing Spin Accumulation and Interfacial Chemical Reconstruction**

Hari Babu Vasili*[†], Matheus Gamino[‡], Jaume Gàzquez*[‡], Florencio Sánchez[‡], Manuel Valvidares[†], Pierluigi Gargiani[†], Eric Pellegrin[†], and Josep Fontcuberta*[‡]

[†]ALBA Synchrotron Light Source, E-08290 Cerdanyola del Vallès, Barcelona, Catalonia, Spain

[‡]Institut de Ciència de Materials de Barcelona (ICMAB-CSIC), Campus UAB, 08193, Bellaterra, Catalonia, Spain

* Author(s) to whom correspondence should be addressed. Electronic mail: hvasili@cells.es (H.B.V.), jgazqueza@gmail.com (J.G.), fontcuberta@icmab.cat (J.F.)

# Supplementary Information SI-1: Structural, Morphological, and Magnetic properties of CoFe₂O₄ thin films and Pt-XAS analysis

## A. X-ray reflectivity and atomic force microscopy

Thicknesses ($t$) of the $CoFe_2O_4$ (CFO) thin films were determined by the x-ray reflectivity (XRR) measurements. Surface morphology was examined using atomic force microscopy (AFM) experiments done in tapping mode.

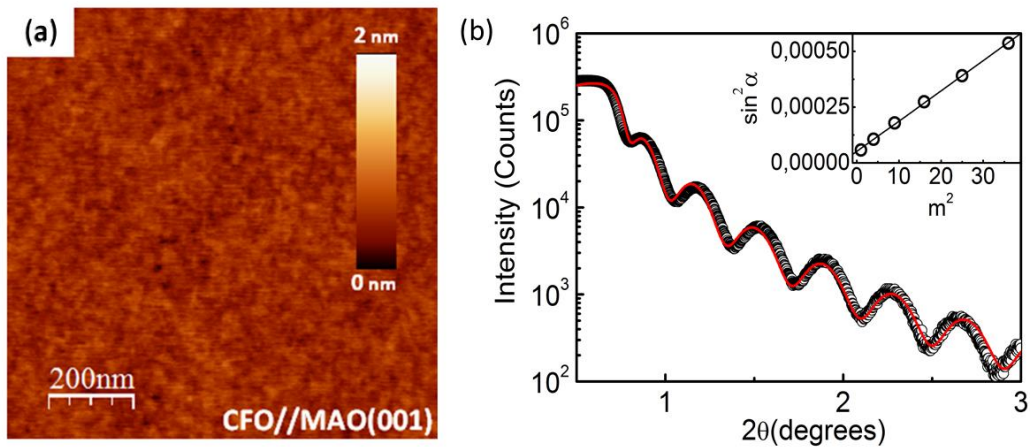

**Figure S1. (a)** *The AFM image of 1µm × 1µm and* **(b)** *The XRR spectra of the CoFe₂O₄ thin film (t ≈ 24 nm). Solid (red) lines across the XRR data indicates best fitting for the thickness calibration, as described in the text. Inset: determination of film thickness from the position ($sin^2$ α) of the Kiesing fringes vs their order (m).*

Figure S1(a) shows the AFM (1 x 1 µm²) image of the CFO thin film ($t$ ≈ 24 nm) grown on MAO(001) substrate. The film exhibits very flat surface with surface roughness ($\sigma$) estimated to be 0.13 nm, *i.e.* less than 1% of its thickness (see the AFM image). In Fig. S1(b) (main panel), we show the XRR data. The red solid line is the best fit obtained using Philips WINGIXA software. The well-defined Kiessig fringes confirm that the surface and interface of the thin film are extremely flat. From the XRR fit, the values of the density, surface roughness, and thickness ($t$) of CFO thin films were found to be ~5.12 g/cm$^3$, 0.17 nm (rms), and 24.1 nm, respectively. The obtained density and the roughness values are in excellent agreement with the theoretical value (5.3 g/cm$^3$) and the AFM value (0.13 nm), respectively.

The thickness determination process is as follows: The thickness is determined from the position ($2\theta$) of the maxima of the reflected x-rays. The positions of the maxima are given by $\sin^2\theta = \sin^2\theta_c + (\lambda/2t)^2 m^2$, where $m$ is the order of the different maxima for the reflected x-rays, and $\lambda$ is the wavelength, $t$ is the thickness of the film, $\theta$ and $\theta_c$ are the reflected and critical angle, respectively. From of the slope of the linear variation of $\sin^2\alpha$ $vs$ $m^2$, we derive a thickness of $\approx$ 24 nm for the CFO film (see the inset of the Fig. S1(b)).

## B. X-ray diffraction: Symmetric reflections and texture

The x-ray diffraction (XRD) patterns in Fig. S2(a) show that the CFO film is fully (00$l$)-oriented without traces of any spurious phase. The XRD reflections are shifted towards smaller angles comparing to the CFO bulk reflections ($c_{bulk}$ = 8.392 Å), indicating that the film has an expanded out-of-plane lattice parameter ($c$-axis) which was determined to be $c_{film}$ = 8.472 Å. This elongation is expected due to the lattice mismatch between the MAO substrate and the CFO film which imposes an epitaxial in-plane compressive strain. The cell parameters of the MAO and CFO bulk are $a_{MAO}$ = 8.083 Å and $a_{CFO}$ = 8.392 Å, respectively. The lattice mismatch, given by $f = \dfrac{[a_{MAO} - a_{CFO}]}{a_{CFO}}$, is $\approx$ -3.68 % (compressive strain).

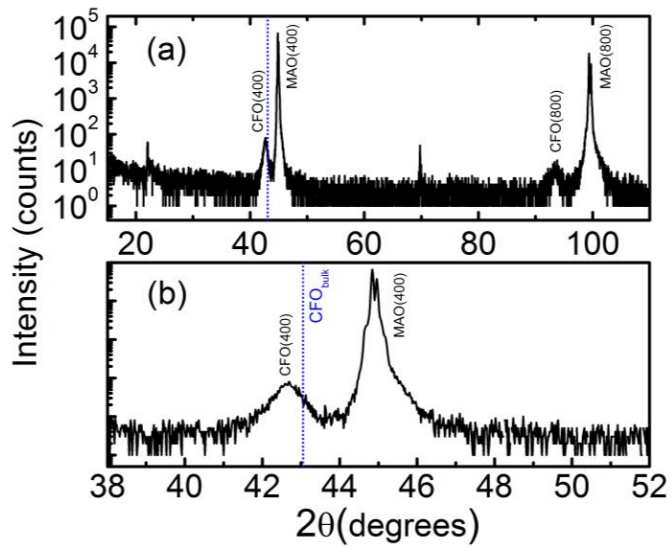

**Figure S2. (a)** *The XRD pattern of 24 nm CoFe₂O₄ thin film* **(b)** *Zoom of the XRD pattern. The blue vertical lines indicate the 2θ position of bulk (004) CoFe₂O₄ reflection.*

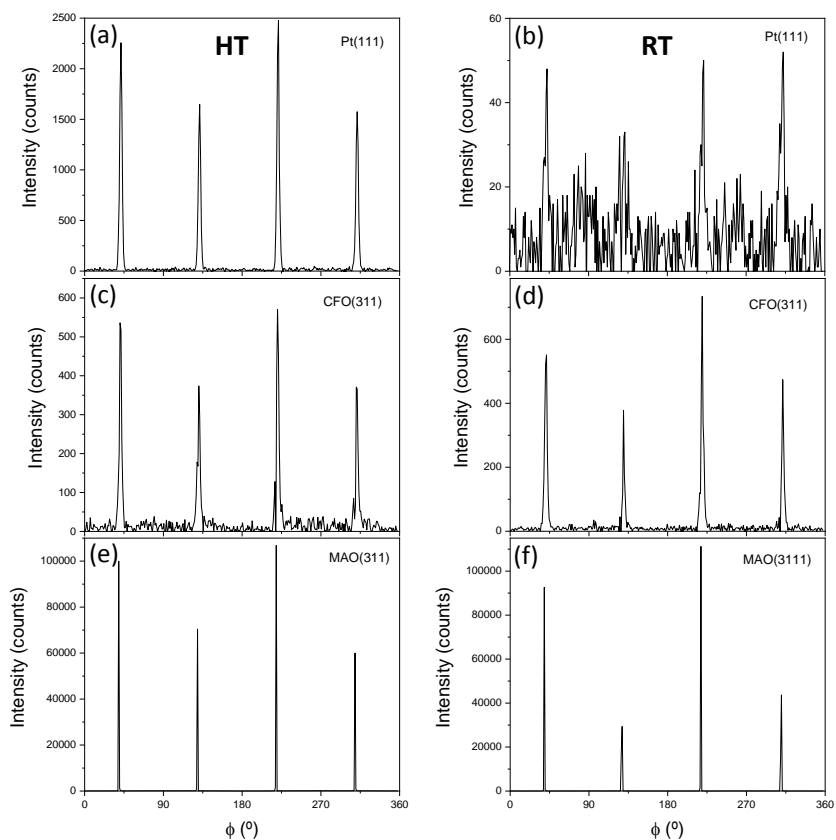

**Figure S3**. *XRD ϕ-scans of (a-b) Pt(111), (c-d) CFO(311), and (e-f) MAO(311) asymmetric reflections corresponding to the sample HT (panels a, c, and d) and sample RT (b, d and f).*

## C.  X-ray diffraction and electron microscopy: Epitaxy

Fig. S3 shows the XRD φ-scans around asymmetrical reflections corresponding to the HT (left panels) and the RT (right panels) samples. The scans were acquired using a Bruker D8 Advance diffractometer with Cu-Kα radiation and area detector. The CFO films, deposited in both samples at same substrate temperature (450 °C), are epitaxial and present cube-on-cube epitaxial relationship with the $MgAl_2O_4$ (MAO) substrate. The Pt(111) φ-scans corresponding to the HT sample shows four peaks confirming the epitaxial growth of Pt at 400 °C, presenting cube-on-cube epitaxial relationship with the bottom CFO film and the MAO substrate. In contrast, the Pt(111) φ-scans of the RT sample shows very low intensity peaks indicating that only very small regions of the Pt deposited at RT are epitaxial.

The above XRD data is in good agreement with the high resolution scanning transmission electron microscopy (STEM) measurements carried out on the HT and RT samples. Fig. S4(a,b) or S5(a,b) show the simultaneously acquired cross-sectional Z-contrast and bright-filed (BF) high resolution images, for HT or RT samples, respectively. The images are viewed along the [010] of MAO zone-axis. The high contrast of the Pt layer in the Z-contrast images restrain the capability of distinguishing the atomic columns of either the CFO layer or the MAO substrate. However, they are clearly visible in the BF images [see Fig. S4(c) or S5(c) for a zoom-in picture of CFO/MAO interface for HT or RT samples, respectively]. The fast Fourier transform (FFT) patterns of the above CFO/MAO interface images show (002) and (200) reflections for both the layer and the substrate [see Fig. S4(d) or S5(d) for HT or RT samples, respectively]. The yellow and red color circles mark the positon of (002) and (200) reflections, respectively. Inset S4(d) or S5(d) shows a magnified image of the (400) reflection, in which the splitting of the CFO layer (in white color) and the MAO substrate (in yellow color) is observed. Overall, the high magnification BF images and the corresponding FFT patterns unambigously confirms a cube on cube epitaxial relationship between the film and the substrate.

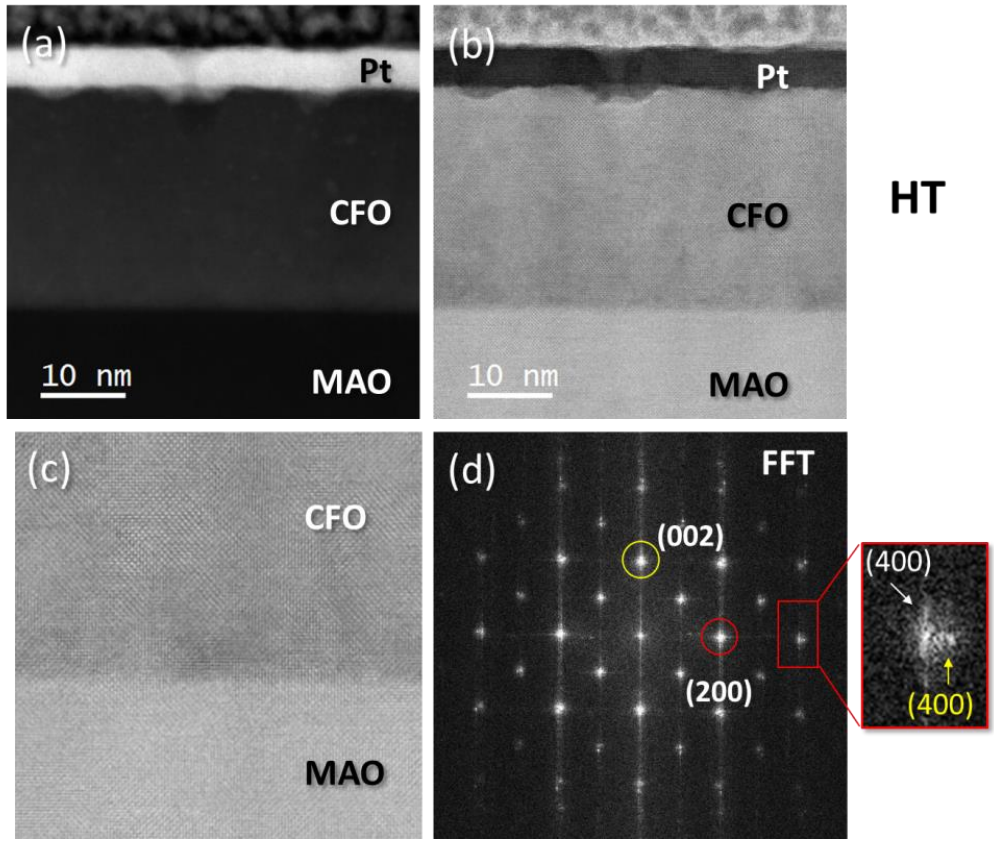

**Figure S4.** *(a) and (b) show the simultaneously acquired cross-sectional Z-contrast and bright filed (BF) high resolution images, respectively, of the HT sample. (c) Zoom-in picture of the BF image of CFO/MAO interface. (d) FFT pattern of the CFO/MAO intercace. Inset (d) shows the magnified image of (400) reflection where a splitting of the CFO layer (in white color) and the MAO substrate (in yellow color) is observed. From both the above (c) BF image and (d) FFT pattern, we confirm a cube on cube epitaxial relationship between the film (CFO) and the substrate (MAO).*

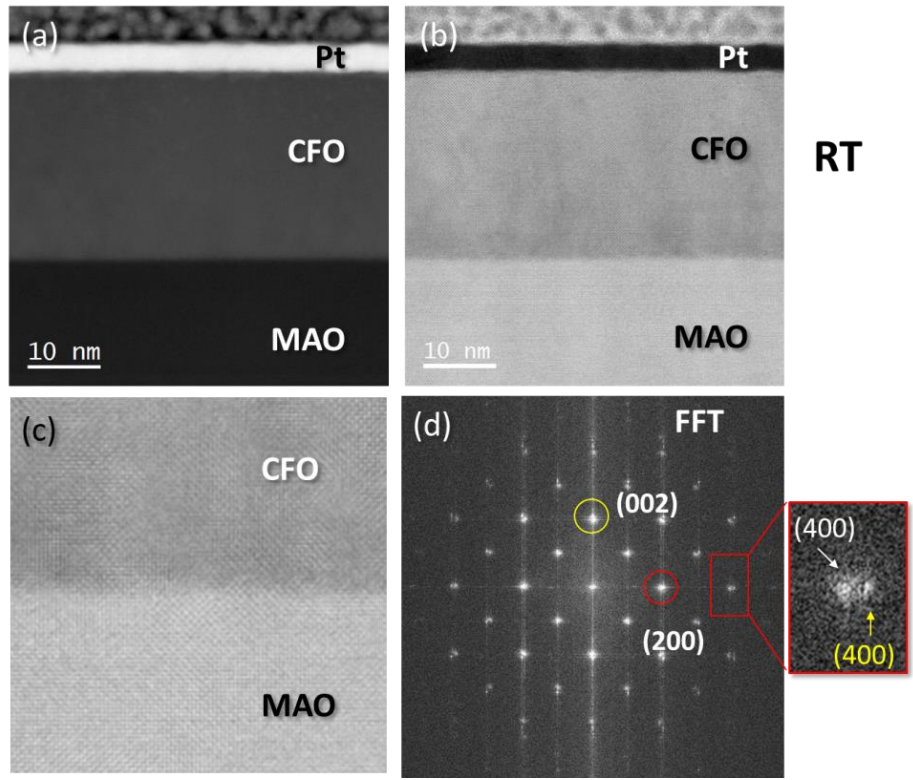

**Figure S5.** *(a) and (b) show the cross-sectional Z-contrast and bright filed (BF) high resolution images, respectively, of the RT sample. (c) Zoom-in picture of the BF image of CFO/MAO interface. (d) FFT pattern of the CFO/MAO intercace. Inset (d) shows the (400) splitting of the CFO and MAO. From both the above (c) BF image and (d) FFT pattern, we confirm the cube on cube epitaxial relationship between the film (CFO) and the substrate (MAO).*

## D. Magnetization

Magnetic properties of the CFO thin films were measured using a superconducting quantum interference device (SQUID) magnetometer (MPMS system of Quantum Design). In the SQUID measurements, the diamagnetic contribution to the susceptibility ($\chi_d$) arises from the MAO substrate and the sample holder. These contributions were evaluated by performing a linear regression of the high field magnetization region (between 65 kOe and 70 kOe) measured with H in-plane. This method assumes that the magnetization of the film is fully saturated in this field region and thus neglects any possible high-field differential susceptibility coming from the film. The average value was found to be $\chi_d = (-1.3\pm0.8) \times 10^{-6}$ emu.Oe$^{-1}$.cm$^{-3}$ and it was subtracted from the magnetization, $M(H)$, collected either $H//[100]$ or $H//[001]$ (see Fig. S6).

The magnetization loops indicate an in-plane easy axis with a coercive field of 6.2 kOe ($H // [100]$) and a large remanent magnetization ($M_r/M_s$ = 50%) (see Fig. S6). The high field in-plane magnetization value is about 335 $emu/cm^3$ (2.65 $\mu_B$/f.u.), which is similar to the reported values of CFO films of similar thickness (see the manuscript for a detailed discussion).

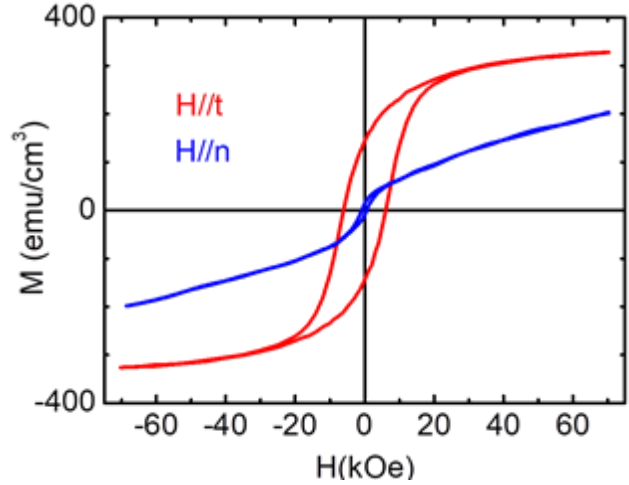

**Figure S6.** *In-plane (**H**||**t**) and out-of-plane (**H**//**n**) magnetization loops of CoFe$_2$O$_4$ (001) thin film (t = 24nm) grown on MAO (001).*

### E. Longitudinal magnetoresistance of the HT Pt/CFO(24 nm)//MAO(001) sample

In Figure S7(a), we show the longitudinal magnetoresistance ($R_L$) $\left[ R_L = \dfrac{R(H) - R(H = 0)}{R(H = 0)} \right]$ measurements of the high temperature, HT (400 ºC) grown Pt/CFO (24 nm)//MAO recorded using magnetic fields **H** applied along three different orientations. In $R_L(\boldsymbol{H}//\boldsymbol{t})$ and $R_L(\boldsymbol{H}//\boldsymbol{n})$ $\boldsymbol{H}$ is orthogonal to the current direction $\boldsymbol{j}$ and in $R_L(\boldsymbol{H}//\boldsymbol{j})$ $\boldsymbol{H}$ is parallel to $\boldsymbol{j}$. We notice that $R_L(\boldsymbol{H}//\boldsymbol{n})$ is negative and has a monotonic field-dependence. In contrast, $R_L(\boldsymbol{H}//\boldsymbol{j})$ and $R_L(\boldsymbol{H}//\boldsymbol{t})$ display a non-monotonic behavior by having a positive and hysteretic magnetoresistance behavior at low fields and a negative magnetoresistance response at high fields (see Figure S7(b)). The non-monotonic response observed for $\boldsymbol{H}//\boldsymbol{j}$ and $\boldsymbol{H}//\boldsymbol{t}$ could be due to the field-dependent domain reorientations of different magnetization components.

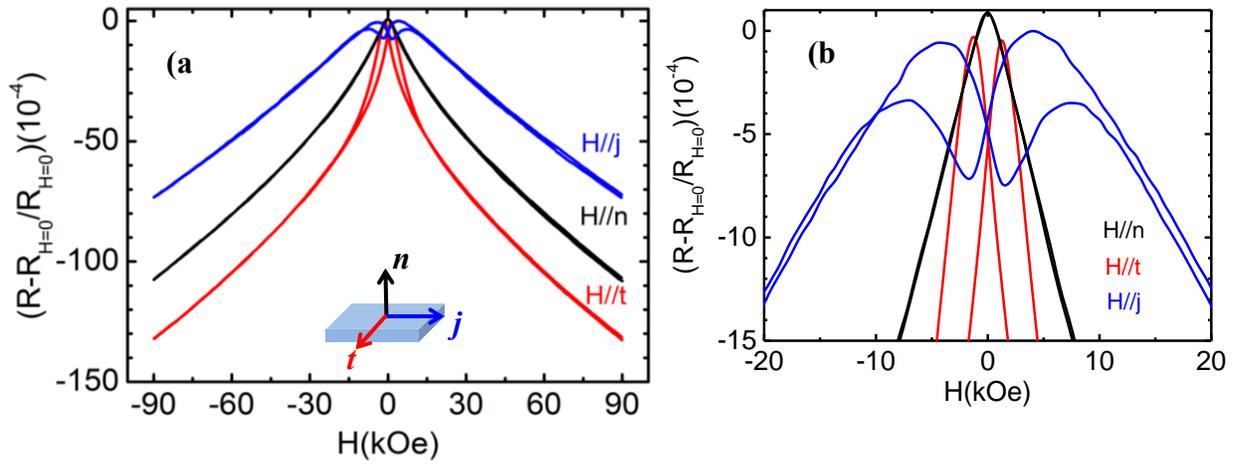

**Figure S7. (a)** *The longitudinal magnetoresistance, $R_L$, of HT Pt/CFO(24 nm)//MAO sample as function of the magnetic field applied along t-, j-, and n-directions.* **(b)** *Zoom of the $R_L$ response at low fields.*

## F.  Transverse magnetoresistance of the HT and RT Pt/CFO(24 nm)//MAO(001) samples

The transverse magnetoresistance ($R_T$) of these HT and RT films has been measured by using a simple four-point contact configuration. Contacts have been hand-made by silver paint. Data have been recorded from +90 kOe to -90 kOe field (H) excursions either at $\boldsymbol{H}//\boldsymbol{n}$ or $\boldsymbol{H}//\boldsymbol{t}$ configurations. Data are depicted in Figure S8 where the results are plotted in panels (b) and (c). For a convenient comparison, in the same plot (panel (a)), we include the longitudinal magnetoresistance data of Fig. 3 in the manuscript. Data in these figures are just the raw data. No attempt has been made to remove the possible longitudinal resistance contribution.

It is obvious in panel (b) that the HT sample displays a large anomalous Hall effect contribution that dominates the $R_T(H)$ behavior. This AHE contribution is clearly absent in the RT sample. As argued in the manuscript the AHE signal comes from the presence of a ferromagnetic-metallic alloy at the Pt/CFO interface. This alloy is also responsible for the larger longitudinal magnetoresistance shown in panel (a).

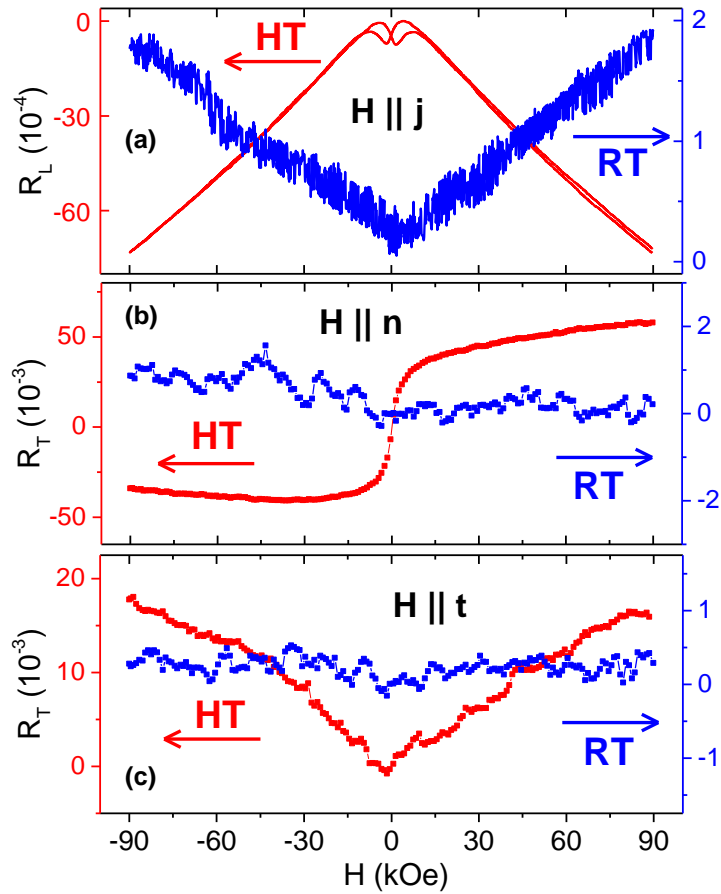

**Figure S8. (a)** *The longitudinal magnetoresistance, $R_L$, of HT and RT Pt/CFO(24 nm)//MAO samples as function of the magnetic field applied along **j (H//j)**. Transverse magnetoresistance, $R_T$, of HT and RT samples as function of the magnetic field applied along:* **(b)** ***H//n*** *and* **(c) H//t.**

**G. Comparison of the x-ray absorption at Pt-M$_3$ edges of the HT and RT Pt(4nm)/CFO(24 nm)//MAO(001)**

In order to compare the white line (WL) intensity of the Pt-M$_3$ XAS of the HT and RT grown Pt/CFO samples, the spectra were normalized to their post edge oscillations for minimizing discrepancy related to the signal background [see Fig. S9(a)]. The WL intensity at the Pt-L$_3$ edge is known to provide a direct hint to the Pt oxide formation in the PtO$_x$ alloys. For instance, the L$_3$ edge has a large intensity for PtO$_{1.6}$ ($\approx$2.2) and exhibit smaller intensities for lowering the oxygen concentration (see Table S1). In our case, we have probed the Pt-M$_3$ edge rather than the Pt-L$_3$. The WL intensities of M$_3$ edges are smaller compared to the L$_3$ edges because of the shorter absorption lengths in the M edges. Therefore, a suitable calibration is needed for comparison. With this aim, we have compared our previously published Pt-M$_3$ data on Co/Pt [Ref. 1] to the recently reported Pt-L$_3$ data on Co/Pt [Ref. 2]. We conclude that the WL intensity ratio M$_3$:L$_3$ of Pt is about 1.1:1.3.

In our work, the WL intensities of the RT and HT samples were found to be 1.2 and 1.12, respectively [Fig. S9(a)]. If these intensities are scaled up to the L$_3$ edges, the RT and HT samples show 1.42 and 1.32, respectively. These values are very much similar to the reported Pt/CFO of Pt grown at room temperature [1.4, Ref. 2] and 400$^0$C [1.25, Ref. 1]. Our data unambiguously reveal that the Pt of HT sample is metallic while the RT sample probably has some slight oxygen adsorption. This is understood from the fact that the Pt oxide phase reduces with the temperature [see Fig. S9(b)]. Therefore, the Pt at the Pt/CFO interface, when Pt is grown Pt at 400$^0$C (HT) becomes fully metallic-like. All in all, our XAS data rules out any possible PtO$_x$ phase formation at the Pt/CFO interface after growing the Pt at 400 °C. Instead, the data prove that the Pt from the HT sample is metallic.

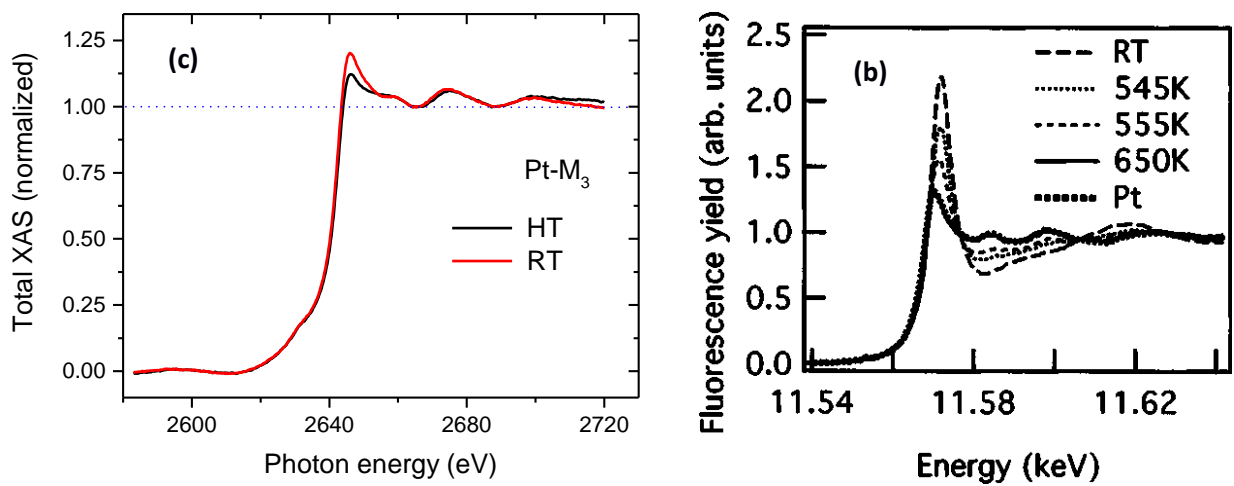

**Figure S9. (a)** *Pt-M₃ XAS of HT and RT grown Pt/CFO(24nm)//MAO(001).* **(b)** *Pt-L₃ XAS as a function of annealing temperatures, adapted from Ref.* [3].

| Edge | Sample | whiteline intensity | Ref. |
|:---:|:---:|:---:|:---|
| **Pt-L₃** | PtO₁.₆ | 2.20 | A. V. Kolobov *et al.* Appl. Phys. Lett. **86**, 121909 (2005) |
| | PtO₁.₃₆ | 1.50 | |
| | Metallic Pt | 1.25 | |
| | Pt/YIG | 1.30 | S. Geprägs *et al.* arXiv:1307.4869v1 |
| | Pt/CFO | 1.24 | M. Valvidares *et al.* Phys. Rev. B **93**, 214415 (2016) |
| | Pt/CFO | 1.40 | M. Collet *et al.* Appl. Phys. Lett. **111**, 202401 (2017) |
| | Co/Pt | 1.30 | |
| **Pt-M₃** | Co/Pt | 1.10 | M. Valvidares *et al.* Phys. Rev. B **93**, 214415 (2016) |

**Table S1.** *Whiteline intensities of some Pt alloys reported mainly at Pt-L₃ edges.*

# Supplementary Information SI-2: Magnetic properties of Pt/CFO (40 nm)//MAO (001) thin film before and after Pt deposition at high temperature

### A. Magnetization

The magnetization of the sample (before and after Pt growth) was measured using the SQUID magnetometer for either in-plane ($H$//) or out-of-plane ($H\perp$) directions to the sample surface. The CFO film (before Pt deposition) exhibits a typical $M$ $vs$ $H$ hysteresis loop (see the left panel of Fig. S10). The film shows an in-plane magnetic anisotropy and a saturation magnetization of $H$// is 272 emu/cm$^3$ (2.23 $\mu_B$). After the Pt (4 nm) growth at 400 °C, the magnetization was found to increase for both the in-plane and out-of-plane directions (see right panel of Fig. S10): the high field magnetization for $H$// is 385 emu/cm$^3$ (3.15 $\mu_B$/atom. The relative change of the magnetization data is in agreement with that observed in the RT and HT Pt/CFO(24 nm)//MAO samples described in the manuscript. Therefore, the growth of Pt at 400 °C, either *ex-situ* or *in-situ*, leads to an enhancement of the magnetization.

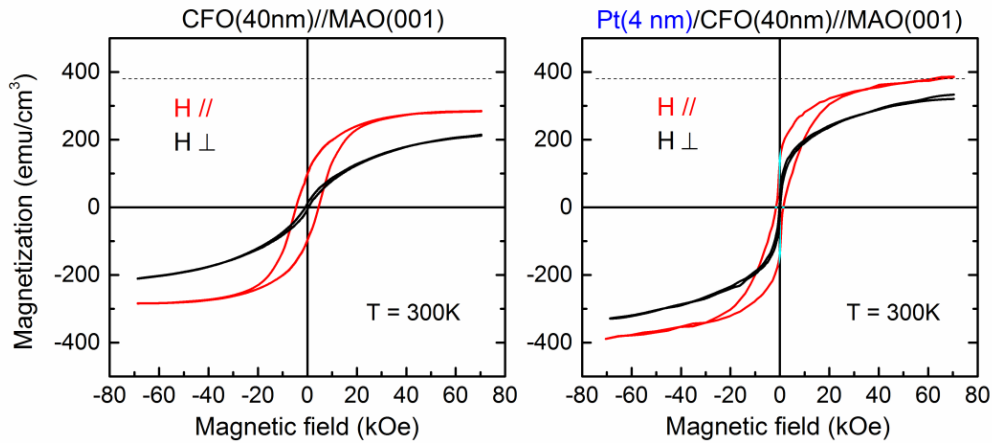

**Figure S10.** *M vs H hysteresis loops of CFO(40 nm)//MAO(001) sample before and after growing the Pt (4 nm) shown in left and right hand side panels, respectively.*

## B. X-ray spectroscopy

We performed XAS and XMCD measurements on this sample after HT Pt growth. The (Fe,Co)-L$_{2,3}$ XAS and XMCD spectral features recorded in TEY mode at grazing incidence (20 degrees from the film plane) at 300 K, are shown in Figure S11. These spectra are completely similar to those observed in the HT Pt(4 nm)/CFO(24 nm)//MAO(001) sample described in the Manuscript (see Figs. 6(c,d)). Data indicate - as discussed in the manuscript – that: (*i*) the iron ions have a mixed contribution of metallic Fe$^0$ and partial Fe$^{3+}$ ions and (*ii*) the cobalt is in mainly metallic Co$^0$ with perceptible traces of Co$^{2+}$ ions. Overall, this spectroscopic data (*ex-situ* Pt growth at 400 °C) is in very well agreement to the *in-situ* grown Pt/CFO sample data.

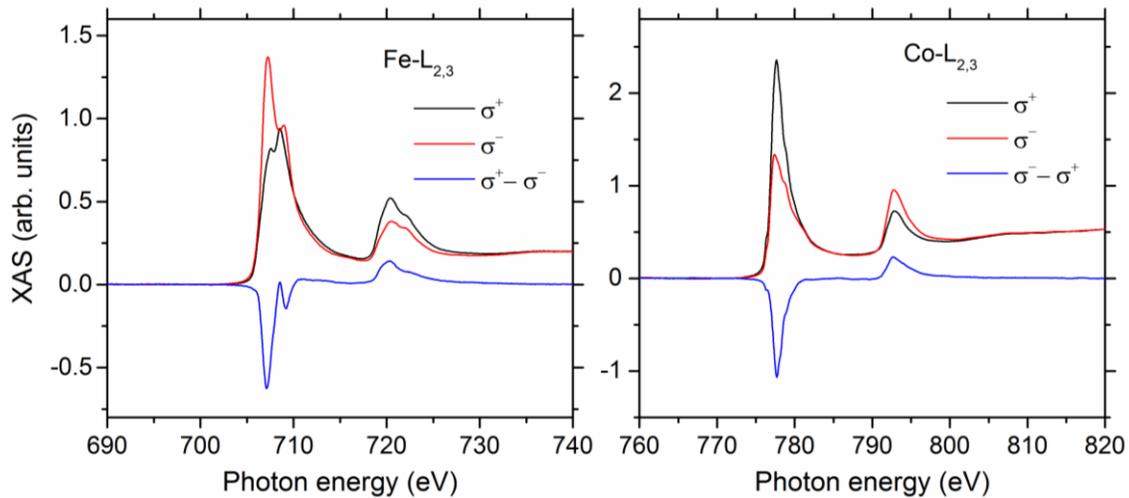

**Figure S11.** *Fe-L$_{2,3}$ and Co-L$_{2,3}$ XAS and XMCD spectra of the Pt(4nm)/CFO(40 nm)//MAO(001) sample shown in left and right hand side panels, respectively. The spectra were recorded at TEY mode at grazing incidence at 300 K.*

**Supplementary Information SI-3**: Sum rules and atomic magnetic moments of HT- and RT-grown Pt/CFO(24 nm)//MAO(001)

We have determined the spin ($m_s$) and orbital ($m_l$) magnetic moments of Fe and Co ions and so the CFO films per formula unit by applying the sum rules to the integrated intensities of total XAS and XMCD spectra.[4-6] A few comments about the evaluation of magnetic moments are relevant. Firstly, the spin sum rule errors, associated to the core-valence exchange interactions,[7] were properly accounted for the calculations because the effective spin moments of $3d^7$($Co^{2+}$) and $3d^5$($Fe^{3+}$) systems have the reported errors of 8% and 31%, respectively. Secondly, we have taken the number of holes for $Co^{2+}$(2.79), $Fe^{3+}$(4.7), $Co^0$(2.49), and $Fe^0$(3.39) from the literature.[5, 8]

As we already discussed in the manuscript [see Fig. 6(a,b)], the RT sample exhibit the conventional XAS of CFO fine structure, therefore, we considered for $Fe^{3+}$ and $Co^{2+}$ the hole numbers given above. In Table S2, we report the $m_s$, $m_l$, and total ($m_{tot}$) magnetic moments of the Co and Fe ions per CFO formula unit (f. u.). The $m_{tot}$ of RT sample was found to be 2.40(5) $\mu_B$/f.u.. The $m_{tot}$ of RT sample is in good agreement with our SQUID magnetization data (2.69 $\mu_B$) and within the range of values reported for CFO thin films (2-3 $\mu_B$) at 300 K.

| Sample | Absorption Edge | $m_s$ /f.u. ($\mu_B$) | $m_l$ /f.u. ($\mu_B$) | $m_{tot} = m_s + m_l$ /f.u. ($\mu_B$) (±0.05) | $\dfrac{m_l}{m_s}$ |
|---|---|---|---|---|---|
| RT | Co-$L$ | 1.03 | 0.46 | 1.49 | 0.45 |
|  | Fe-$L$ | 0.88 | 0.03 | 0.91 | 0.03 |
|  |  |  |  | **2.40** |  |

| Sample | Absorption Edge | $m_s$ /atom ($\mu_B$) | $m_l$ /atom ($\mu_B$) | $m_{tot} = m_s + m_l$ /atom ($\mu_B$) (±0.05) | $\dfrac{m_l}{m_s}$ |
|---|---|---|---|---|---|
| HT | Co-$L$ | 1.19 | 0.19 | 1.38 | 0.16 |
|  | Fe-$L$ | 1.86 | 0.12 | 1.98 | 0.06 |
|  | Pt-$M$ | 0.24 | unknown | 0.24 | n.a. |
|  |  |  |  | **3.60** |  |

**Table S2.** *Spin ($m_s$), orbital ($m_l$), and total ($m_{tot}$) magnetic moments of the (i) Co and Fe ions RT sample (per CFO formula unit, f.u.)* (**Upper table**) *and (ii) Co and Fe atoms in HT sample* (**Lower table**) *calculated using the XMCD sum rules on Co- and Fe-L absorption edges.*

As the HT sample exhibit mainly the metallic Fe/Co in the XMCD spectra [see Figure 6(c,d) in the manuscript], we have removed the persisting traces of $Fe^{3+}/Co^{2+}$ ions from both the XAS and XMCD spectra of HT sample in order to determine the atomic magnetic moments of Fe/Co atoms in the Pt-(Fe,Co) intermixing layer. After removing ~30 % of $Fe^{3+}$ and 5 % of $Co^{2+}$ XAS of RT sample signals from the above HT sample signal, the spectra appears very similar to those observed for metallic Fe/Co (see the Figure S12). We have employed the XMCD sum rules on these metallic only Fe/Co spectra. It turns out that the Fe exhibit a bulk-like moment of ~2 $\mu_B$/atom while the Co moment is little shorter than its bulk moment (1.7 $\mu_B$ ). Overall, the total $m_{tot}$ of the Pt-(Co,Fe) alloy is 3.60(5) $\mu_B$ including the Pt-$M_3$ magnetic moment of 0.24 $\mu_B$. In any case, the above intermixed alloy exhibit significantly larger $m_{tot}$ than the RT sample.

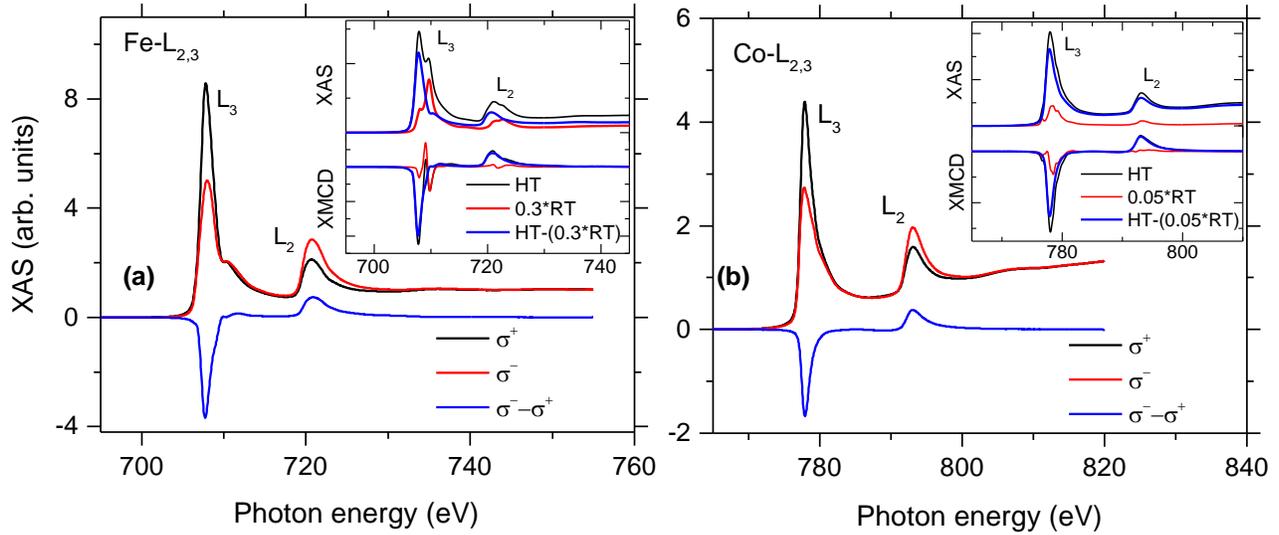

**Figure S12. (a,b)** *Fe-$L_{2,3}$ and Co-$L_{2,3}$ XAS and XMCD spectra of the HT grown Pt/CFO//MAO(001) sample after removing the $Fe^{3+}$and $Co^{2+}$ scaled intensities of 30 % and 5 % of the RT signals, respectively. The insets show the extraction of the metallic only $Fe^0$ and $Co^0$ XAS ($\sigma^+$) and XMCD ($\sigma^+$$-$$\sigma^+$).*